\documentclass[10pt,conference]{IEEEtran}
\usepackage{listings}

\usepackage{xcolor}
\usepackage{color}
\definecolor{mygreen}{rgb}{0,0.6,0}
\definecolor{mymauve}{rgb}{0.58,0,0.82}

\usepackage{colortbl}
\definecolor{mygray}{gray}{.85}

\ifCLASSINFOpdf
\else
\fi
\usepackage{hyperref}
\usepackage{url}
\usepackage{graphicx}
\usepackage[noend]{algpseudocode}
\usepackage{algorithmicx,algorithm}
\usepackage{amsmath,amssymb,amsfonts}

\usepackage{enumitem}
\usepackage{listings}
\usepackage{mdframed}
\usepackage{multirow}
\usepackage{balance}

\usepackage{subfig}
\usepackage{tikz}
\usetikzlibrary{automata,arrows,positioning}

\usepackage{xspace}

\newcommand{\tool}{{\sc LJS}\xspace}
\newcommand{\supp}{{\sf supp}}

\begin{document}

\title{
	Model-based Automated Testing of JavaScript Web Applications via Longer Test Sequences}

\author{\IEEEauthorblockN{Pengfei Gao and Fu Song}
\IEEEauthorblockA{School of Information Science and Technology\\
ShanghaiTech University}
\and
\IEEEauthorblockN{Taolue Chen}
\IEEEauthorblockA{School of computer Science and Engineering\\ Nanyang Technological University}
\and
\IEEEauthorblockN{Yao Zeng}
\IEEEauthorblockA{School of Computer Science and Software Engineering\\ East China Normal University}
\and
\IEEEauthorblockN{Ting Su}
\IEEEauthorblockA{School of computer Science and Engineering\\ Nanyang Technological University}
}






\maketitle

\begin{abstract}
JavaScript has become one of the most widely used languages for Web development. However, it is challenging to ensure the correctness and reliability of  Web applications written in JavaScript, due to their dynamic and event-driven features. A variety of dynamic analysis techniques for JavaScript Web applications have been proposed, but they are limited in either coverage or scalability. In this paper, we propose a model-based automated approach to achieve high code coverage in a reasonable amount of time via testing with longer event sequences. We implement our approach as the tool \tool, and perform extensive experiments on 21 publicly available benchmarks (18,559 lines of code in total). On average, \tool achieves 86.4\% line coverage in 10 minutes, which is 5.4\% higher than that of {\sc JSDep}, a breadth-first search based automated testing tool enriched with partial order reduction. In particular, on large applications, the coverage of \tool is 11-18\% higher than that of {\sc JSDep}. Our empirical finding supports that longer test sequences can achieve higher code coverage in JavsScript testing.
\end{abstract}

\section{Introduction}
JavaScript is a highly dynamic programming language with first-class functions and ``no crash" philosophy,
which allows developers to write code without type annotations, and to generate and load code at runtime. Partially due to these programming flexibilities,
Web applications based on JavaScript are gaining increasing popularity. These features are, however, double-edged sword, making these Web applications 
prone to errors and intractable to static analysis.

Dynamic analysis has proven to be an effective way to test JavaScript Web applications
~\cite{SAHMMS10,artzi2011framework,Mesbah2012Crawling,MM12,HT12,Gibbs2013Jalangi,Mirshokraie2013Efficient,FM13,MMP13,Mirshokraie14,FMM14,li2014symjs,Pradel2014EventBreak,FW15,MP15,SNGC15,TULG15,MP16,DRS16,saner16,AGMPSSS17}.
Since it requires testcases to explore the state space of the application, various
approaches for automated testcase generation have been developed in literature,
which can generate event sequences and/or associated input data of events.
The event sequences concern the order in which event handlers are executed (e.g., the order
of clicking buttons), while the input data concerns the choice of values (e.g., strings, numbers and objects).
The generation of both event sequences and input data is important to achieve a high code coverage, and has been extensively studied.

In general, event sequences are generated by randomly selecting event handlers with heuristic search strategies~\cite{SAHMMS10,artzi2011framework,Mesbah2012Crawling,Mirshokraie14,Pradel2014EventBreak}.
These approaches are able to analyze large real-life applications, but are usually left with a low code coverage.
One possible reason, as mentioned before~\cite{li2014symjs,sung2016static}, is that the event sequences are insufficiently long to explore parts of the code which may trigger the error.
For example, in an experiment of~\cite{li2014symjs}, the uncovered code of the benchmark \emph{tetris} is mainly due to the function \emph{gameOver} which will only be invoked after a long event sequence. %
For traditional white-box testing and GUI testing, it has been shown that increasing the length of test cases could improve coverage and failure detection~\cite{Arc10a,AGWX08,FG09,XM06,CA15}. However, this has not been fully exploited in testing JavaScript Web applications. One of the reasons is that existing approaches usually generate event sequences from scratch by iteratively
appending events to the constructed sequences up to a maximum bound,
and the number of event sequences may blow up exponentially in terms of this bound. Therefore, for efficiency consideration, the maximum bound  often have to be small (for instance, less than 6 \cite{li2014symjs}).
%
To mitigate these issues, pruning techniques (e.g. mutation testing~\cite{Mirshokraie2013Efficient,MP15} and partial order reduction~\cite{sung2016static}) were proposed to remove redundant event sequence, which allow to explore \emph{limited} longer event sequences in a reasonable time.
On the other hand, the input data is generated by either randomly choosing values with lightweight heuristic strategies~\cite{artzi2011framework,Mesbah2012Crawling,HT12},
or using heavyweight techniques (e.g., symbolic/concolic execution)~\cite{SAHMMS10,Gibbs2013Jalangi,FMM14,li2014symjs,SNGC15,TULG15,DRS16}.
These works either consider unit testing or 
usually simply reuse the aforementioned methods to generate event sequences.


In this work, we focus on the event sequence generation. In particular, we propose
a novel model-based automated testing approach to achieve a high code coverage in a reasonable time
by generating and executing long event sequences.
Our approach mainly consists of two key components: model constructor
and event sequence generator. The model constructor iteratively queries an execution engine to generate a finite-state machine (FSM) model. 
It explores the state space
using long event sequences and
in a way to avoid prefix event subsequences re-executing and backtracking.
To improve scalability, we propose a state abstraction approach, as well as a weighted event selection strategy, to construct small-sized FSM models. The event sequence generator creates long event sequences by randomly traversing the FSM model from the initial state. 
We implement our approach in a tool \textbf{L}onger \textbf{J}ava\textbf{S}cript (\tool).
To compare with other methods, we also implemented
a random event selection strategy, and event sequence generator from {\sc JSDep}~\cite{sung2016static} based on the FSM model. 

In summary, the contributions of this paper are 
\begin{itemize}
	\item A new model-based automated testing approach for client-side JavaScript Web applications via longer event sequences;
	\item An implementation of our approach as a tool \tool; and
	\item An evaluation of 21 benchmarks from {\sc JSDep}~\cite{sung2016static} including 18,559 lines of code in total.
\end{itemize}

\tool is available from \url{https://github.com/JavaScriptTesting/LJS}.
On average, it achieves 86.4\% line coverage in 10 minutes
and is 5.4\% higher than that of the tool {\sc JSDep}~\cite{sung2016static}.
On large applications, the coverage of \tool is 11-18\% higher than that of {\sc JSDep}.
We have found that long event sequences can indeed improve the coverage with respect to the application under test. This paper then provides concrete, empirically validated approaches to generate long event sequences.

\medskip
\noindent\emph{Structure.} Section~\ref{sec:pre} introduces preliminaries,  a running example and an overview of \tool.
Section~\ref{secl:mothod} describes the methodology of \tool.
Section~\ref{sec:exp} presents experimental results.
Section~\ref{sec:rel} discusses related work.
We conclude in Section~\ref{sec:conc}.

\section{Preliminaries and Overview}\label{sec:pre}

\subsection{Preliminaries}
\subsubsection{JavaScript Web Application}
A client-side Web application, which is the main focus of the paper, consists of HTML/Script files which
are executed by a Web browser. When a browser loads a Web page, it parses the HTML/Script files, represents them
as a Document Object Model (DOM) tree, and then executes the top-level script code.
Each node in the DOM tree represents an object on the Web page and
may also be associated with a set of events.
Each event may have some event handlers (e.g., callback functions such as \emph{onload} and \emph{onclick}) which are either statically registered inside the HTML file
or dynamically registered by executing functions.
When an event occurs (e.g., a button is clicked), the corresponding event handlers are executed sequentially.
Although the browser ensures that each callback function is executed atomically,
the execution of the entire Web application exhibits nondeterminism due to the interleaving of
executions of multiple callback functions.

\subsubsection{DOM Event Dependency}
Given a JavaScript Web application, let $R_c$ and $R_d$ respectively denote the \emph{control} and \emph{data} dependency relation over functions of the application. Formally, for each pair of functions (usually event handlers) $m_1$ and $m_2$, $(m_1,m_2)\in R_c$ (resp. $(m_1,m_2)\in R_d$)
if there are two statements $st_1$ in $m_1$ and $st_2$ in $m_2$ that $st_1$ affects the control (resp. data) flow of $st_2$.
Given two DOM events $e_1$ and $e_2$, $e_2$ is \emph{dependent on} $e_1$, denoted by $e_1\rightarrow e_2$,  
if (1) there are event handlers $m_1$ and $m_2$ of $e_1$ and $e_2$ respectively
such that $(m_1,m_2)\in (R_c\cup R_d)^*$, or (2) the execution of $m_1$ registers, removes, or modifies $m_2$. Two event sequences $\rho_1$ and $\rho_2$ are \emph{equivalent} if
$\rho_1$ can be transformed from $\rho_2$ by
repeatedly swapping adjacent and independent events of $\rho_2$.

\subsubsection{Finite State Machine}
A (nondeterministic) Finite State Machine (FSM) is a tuple $M = (S,I, \delta, s_0)$,
where $S$ is a finite set of states with $s_0\in S$ as the initial state,
$I$ is a finite input alphabet,
$\delta\subseteq S \times I \times S$ is the transition relation.
A transition
$(s_1,e,s_2)\in \delta$ denotes that after reading the input
symbol $e$ at the state $s_1$, the FSM $M$ can move from the state $s_1$ to the state $s_2$.
We denote by $\supp(s)$ the set $\{(e,s')\in I\times S\mid (s,e,s')\in\delta\}$.
Given a word $e_1\cdots e_n\in I^*$, a \emph{run} of $M$ on $e_1\cdots e_n$ is a sequence of
states $s_0s_1\cdots s_n$ such that for every $1\leq i\leq n$,
$(s_{i-1},e_i,s_i)\in\delta$. We denote by $\epsilon$ the empty word.
Note that, in this work, we will use an FSM to represent the behaviors of a JavaScript Web application.
Because of this, we do not introduce the final states as in classic FSMs.
%
%

\begin{figure}[t]
\begin{lstlisting}
<!DOCTYPE html>
<html>
<head>
    <p> Example with 3 checkboxes and 1 button </P>
</head>
<body>
<div id="checkboxes">
  <input id="A" type="checkbox" onclick="FA(this)"> A
  <input id="B" type="checkbox" onclick="FB(this)"> B
  <input id="C" type="checkbox" onclick="FC(this)"> C
</div>
<button id="Submit" type="button" > Submit </button>
<script>
    var count = 0;
    function FA(node) {
        if(node.checked == false) count = count - 1;
        else count = count + 1;
        CheckedEnough();  }
    function FB(node) {
        if(node.checked == false) count = count - 1;
        else  count = count + 1;
        CheckedEnough();  }
    function FC(node){
        if(node.checked == false) count = count - 1;
        else count = count + 1;
        CheckedEnough();  }
    function CheckedEnough() {
        var  b = document.getElementById("Submit");
        if(count >= 3) b.onclick = FSubmit;
        else b.onclick = null;  }
    function FSubmit() {alert("Submit successfully");}
</script>
</body>
</html>
\end{lstlisting}
\vspace*{-0.5em}
\caption{Example HTML page and associated JavaScript code.}
\vspace*{-1.5em}
\label{Figure:1}
\end{figure}

\subsection{Running Example} \label{runningexample}

\begin{figure*}
\scriptsize{
\begin{tikzpicture}[->,thick]
\tikzstyle{level 1}=[sibling distance=60mm,level distance=7mm]
\tikzstyle{level 2}=[sibling distance=20mm,level distance=8mm]
\tikzstyle{level 3}=[sibling distance=6.5mm,level distance=10mm]
\tikzstyle{level 4}=[sibling distance=1.6mm,level distance=12mm]
\node (n0) {$\bullet$}
  child {  node (n00) {$\bullet$}
      child {node (C1)  {$\bullet$}
          child { node  {$\bullet$}
              child { node {$\bullet$}  }
              child { node {$\bullet$}  }
              child { node  {$\bullet$}  }
              edge from parent node[left] {A}
          }
          child {node {$\bullet$}
              child { node  {$\bullet$}  }
              child { node  {$\bullet$}  }
              child { node {$\bullet$}  }
              edge from parent node[left] {B}
          }
          child {node (D1) {$\bullet$}
              child { node  {$\bullet$}  }
              child { node  {$\bullet$}  }
              child { node {$\bullet$}  }
              edge from parent node[left] {C}
          }
          edge from parent node[above] {A}
      }
      child {node (C2)  {$\bullet$}
          child { node  {$\bullet$}
              child { node {$\bullet$}  }
              child { node  {$\bullet$}  }
              child { node  {$\bullet$}  }
              edge from parent node[left] {A}
          }
          child {node {$\bullet$}
              child { node  {$\bullet$}  }
              child { node {$\bullet$}  }
              child { node  {$\bullet$}  }
              edge from parent node[left] {B}
          }
          child {node (D2) {$\bullet$}
              child { node  {$\bullet$}  }
              child { node  {$\bullet$}  }
              child { node {$\bullet$}  }
              child { node[red]  {$\bullet$} edge from parent[red] }
              edge from parent node[left] {C}
          }
          edge from parent node[left] {B}
      }
      child {node (C3)  {$\bullet$}
          child { node  {$\bullet$}
              child { node {$\bullet$}  }
              child { node {$\bullet$}  }
              child { node  {$\bullet$}  }
              edge from parent node[left] {A}
          }
          child {node {$\bullet$}
              child { node  {$\bullet$}  }
              child { node  {$\bullet$}  }
              child { node {$\bullet$}  }
              child { node[red]  {$\bullet$} edge from parent[red] }
              edge from parent node[left] {B}
          }
          child {node (D3) {$\bullet$}
              child { node  {$\bullet$}  }
              child { node {$\bullet$}  }
              child { node  {$\bullet$}  }
              edge from parent node[left] {C}
          }
          edge from parent node[above] {C}
      }
      edge from parent node[above] {A}
  }
  child { node (n01) {$\bullet$}
      child {node (C4)  {$\bullet$}
          child { node  {$\bullet$}
              child { node {$\bullet$}  }
              child { node {$\bullet$}  }
              child { node  {$\bullet$}  }
              edge from parent node[left] {A}
          }
          child {node {$\bullet$}
              child { node  {$\bullet$}  }
              child { node  {$\bullet$}  }
              child { node {$\bullet$}  }
              edge from parent node[left] {B}
          }
          child {node (D4) {$\bullet$}
              child { node  {$\bullet$}  }
              child { node  {$\bullet$}  }
              child { node {$\bullet$}  }
              child { node[red]  {$\bullet$} edge from parent[red] }
              edge from parent node[left] {C}
          }
          edge from parent node[above] {A}
      }
      child {node (C5)  {$\bullet$}
          child { node  {$\bullet$}
              child { node {$\bullet$}  }
              child { node {$\bullet$}  }
              child { node  {$\bullet$}  }
              edge from parent node[left] {A}
          }
          child {node {$\bullet$}
              child { node  {$\bullet$}  }
              child { node  {$\bullet$}  }
              child { node {$\bullet$}  }
              edge from parent node[left] {B}
          }
          child {node (D5) {$\bullet$}
              child { node  {$\bullet$}  }
              child { node  {$\bullet$}  }
              child { node {$\bullet$}  }
              edge from parent node[left] {C}
          }
          edge from parent node[left] {B}
      }
      child {node  (C6) {$\bullet$}
          child { node  {$\bullet$}
              child { node {$\bullet$}  }
              child { node {$\bullet$}  }
              child { node  {$\bullet$}  }
              child { node[red]  {$\bullet$} edge from parent[red] }
              edge from parent node[left] {A}
          }
          child {node {$\bullet$}
              child { node  {$\bullet$}  }
              child { node  {$\bullet$}  }
              child { node {$\bullet$}  }
              edge from parent node[left] {B}
          }
          child {node (D6) {$\bullet$}
              child { node  {$\bullet$}  }
              child { node  {$\bullet$}  }
              child { node {$\bullet$}  }
              edge from parent node[left] {C}
          }
          edge from parent node[above] {C}
      }
      edge from parent node[left] {B}
    }
    child { node (n02) {$\bullet$}
      child {node (C7)  {$\bullet$}
          child { node  {$\bullet$}
              child { node {$\bullet$}  }
              child { node {$\bullet$}  }
              child { node  {$\bullet$}  }
              edge from parent node[left] {A}
          }
          child {node {$\bullet$}
              child { node  {$\bullet$}  }
              child { node  {$\bullet$}  }
              child { node {$\bullet$}  }
              child { node[red]  {$\bullet$} edge from parent[red] }
              edge from parent node[left] {B}
          }
          child {node (D7) {$\bullet$}
              child { node  {$\bullet$}  }
              child { node {$\bullet$}  }
              child { node  {$\bullet$}  }
              edge from parent node[left] {C}
          }
          edge from parent node[above] {A}
      }
      child {node (C8)  {$\bullet$}
          child { node  {$\bullet$}
              child { node {$\bullet$}  }
              child { node {$\bullet$}  }
              child { node  {$\bullet$}  }
              child { node[red]  {$\bullet$} edge from parent[red] }
              edge from parent node[left] {A}
          }
          child {node {$\bullet$}
              child { node  {$\bullet$}  }
              child { node  {$\bullet$}  }
              child { node {$\bullet$}  }
              edge from parent node[left] {B}
          }
          child {node (D8) {$\bullet$}
              child { node  {$\bullet$}  }
              child { node  {$\bullet$}  }
              child { node {$\bullet$}  }
              edge from parent node[left] {C}
          }
          edge from parent node[left] {B}
      }
      child {node (C9)  {$\bullet$}
          child { node  {$\bullet$}
              child { node {$\bullet$}  }
              child { node {$\bullet$}  }
              child { node  {$\bullet$}  }
              edge from parent node[left] {A}
          }
          child {node {$\bullet$}
              child { node  {$\bullet$}  }
              child { node  {$\bullet$}  }
              child { node {$\bullet$}  }
              edge from parent node[left] {B}
          }
          child {node (D9) {$\bullet$}
              child { node  {$\bullet$}  }
              child { node  {$\bullet$}  }
              child { node {$\bullet$}  }
              edge from parent node[left] {C}
          }
          edge from parent node[above] {C}
      }
      edge from parent node[above] {C}
  } ;
\end{tikzpicture}}
\caption{The search tree of our running example, where the missed labels of edges in black color are respectively {\tt A,B,C}, and
the missed labels of edges in red color are {\tt Submit}.\label{fig:searchtree}}
  \vspace{-1em}
\end{figure*}
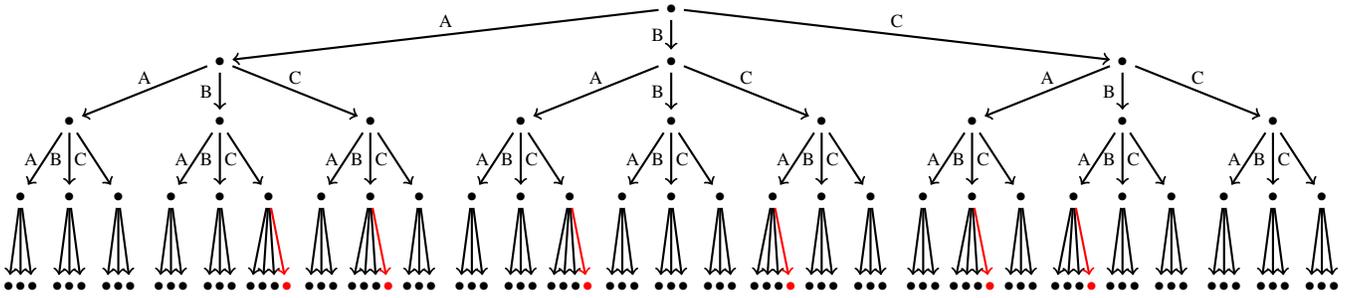

Consider the running example in Figure~\ref{Figure:1}, where
the HTML code defines the DOM elements of three checkboxes and one button. The JavaScript code defines a global variable {\tt count} and five functions manipulating {\tt count} and the DOM elements.
Initially, the three checkboxes, named {\tt A-C}, are unchecked; the button, named {\tt Submit}, does not have any \emph{onclick} event handler. The \emph{onclick} event handlers of the three checkboxes are the functions
{\tt FA-FC} respectively. For each {\tt X}$\in$\{{\tt A,B,C}\},
when the checkbox {\tt X} is clicked (i.e., the corresponding event occurs),
its state (checked/unchecked) is switched, and 
then the \emph{onclick} event handler, namely the function {\tt FX}, is executed.
{\tt FX} first determines whether the state of {\tt X} is checked or not.
If it is checked, then the global variable {\tt count} increases by one,
otherwise {\tt count} decreases by one. Finally,  
{\tt FX} invokes
the function {\tt CheckedEnough} to verify whether the value of {\tt count} is no less than $3$, i.e.,
there are at least three checkboxes in the checked state. If {\tt count}$>=$3, then  the {\tt FSubmit} function
is registered to the \emph{onclick} event handler of the
button {\tt Submit}; otherwise (i.e., {\tt count}$<$3), the \emph{onclick} event handler of the
button {\tt Submit} is removed. If the button {\tt Submit} is clicked when {\tt FSubmit} serves the \emph{onclick} event handler of {\tt Submit},
{\tt FSubmit} gets executed and
prints a message to the console.

In this example, for  {\tt X,Y}$\in$\{{\tt A,B,C}\},
the \emph{onclick} event of {\tt X} is dependent upon the \emph{onclick} event of {\tt Y},
and the \emph{onclick} event of {\tt Submit} is dependent upon the \emph{onclick} event of {\tt X}.

\subsection{Limitations of Existing Approaches}
We now demonstrate why long event sequences are important for automated testing of JavaScript Web applications.
Tools like {\sc Artemis}~\cite{artzi2011framework}
and {\sc JSDep}~\cite{sung2016static} generate event sequences by systematically triggering various DOM events up to a fixed depth.
After loading the Web page, these tools start by exploring all the available events at an initial state.
If a new state is reached by executing a sequence of events, then all the available events at the new state
are appended to the end of the event sequence. The procedure is repeated until time out or a fixed depth is reached. The search tree of our running example up to depth four is depicted in Figure~\ref{fig:searchtree},
where the unshown labels of edges in black  are respectively {\tt A,B,C}, and
those in red are {\tt Submit}.
Each edge labeled by {\tt X}$\in$\{{\tt A,B,C,Submit}\} denotes the execution of the \emph{onclick} event handler of {\tt X},
and each node denotes a state.
For each event sequence $\rho$ of length three, 
if $\rho$ is a permutation of {\tt A;B;C},
then there are four available events {\tt A,B,C,Submit} after $\rho$, otherwise
there are three available events {\tt A,B,C}.

A na\"{\i}ve algorithm (like the default algorithm in {\sc Artemis}) would inefficiently explore the event space, i.e., the full tree in Figure~\ref{fig:searchtree}, and may generate  $6+\sum_{i=1}^4 3^i=126$ event sequences. However, many of them are redundant. 
For instance, the sequences {\tt A;B;C;Submit} and {\tt B;A;C;Submit} actually address the same part of the code, hence one of them is unnecessary for testing purposes. To remedy this issue,
{\sc JSDep} implemented a partial-order reduction in {\sc Artemis} which prunes redundant event sequences by leveraging  DOM event dependencies.

To cover all code of the running example, each event handler of all three checkboxes
has to be executed at least two times (examining checked/unchecked states). Therefore,
a sequence of length seven (e.g., {\tt A;B;C;Submit;A;B;C}) is sufficient to fully cover all the code.
However, if one sets the depth bound of the test sequence to be seven for the full code coverage,
the default search algorithm in {\sc Artemis} and {\sc JSDep} may explore at least
$\sum_{i=1}^73^i=3279$ event sequences. Notice that both {\sc Artemis} and {\sc JSDep} may re-execute previously executed test sequences in order to explore
further the event space, which is time-consuming. Partially because of this, within a time limit these tools often generate and execute only short event sequences. Similar to classic program analysis, it is not hard to envision that short event sequences would hamper the coverage of code.
Indeed in our running example, covering the function {\tt Submit} requires test sequences with length at least 4. Unfortunately, existing approaches suffer from 
the ``test sequence explosion" problem when increasing the depth bound of testing sequences. In this work, we propose a model-based, automated testing approach for JavaScript Web applications, aiming to
generate long event sequences to improve the code coverage, but
do so in a clever way to mitigate the issue of exponential blowup. 

\subsection{Overview of Our Approach}

\begin{figure}
  \centering
  \includegraphics[width=0.48\textwidth]{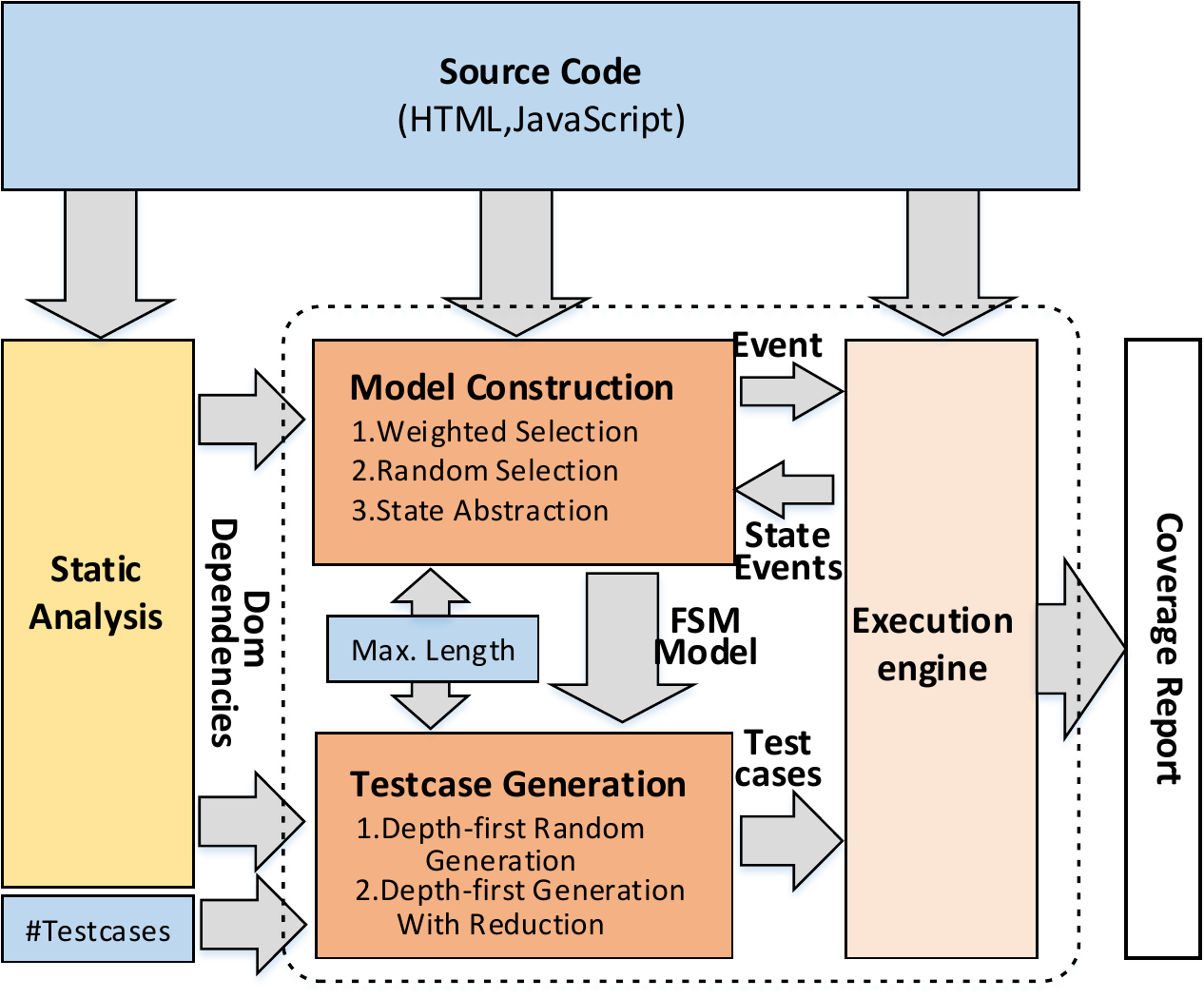}
  \caption{Framework overview of \tool.}\label{Figure:2}
  \vspace{-1em}
\end{figure}

Figure~\ref{Figure:2} presents an overview of our approach implemented in \tool, which consists of four components: static analysis,  execution engine, model construction and testcase generation.
Given the HTML/JavaScript source file(s) of a (client-side JavaScript Web) application,
a  length bound of event sequences (Max.~Length), and a bound of the number of event sequences to be generated (\#Testcases) as inputs, \tool outputs a (line) coverage report.
Internally, \tool goes through the following steps:
(1) compute the DOM event dependencies via static analysis.
(2) construct an FSM model of the application with a variant of depth-first search by leveraging an execution engine and the DOM event dependencies. The FSM model is used to generate long event sequences.
(3) all the generated event sequences are iteratively executed on the execution engine and output the  coverage report. We now elaborate these steps in more detail.

\smallskip
\noindent
{\bf Static Analysis}. Given the HTML/JavaScript source file(s) of an application,
this component computes the DOM event dependencies. In our implementation,
we leverage {\sc JSdep}~\cite{sung2016static} to compute DOM dependency.
It first constructs a control flow graph (CFG) of the JavaScript code, and
then traverses the CFG and encodes the control and data flows (i.e. dependency relations) in Datalog. 
The DOM event dependencies are finally computed via a Datalog inference
engine. 
More details can be found in~\cite{sung2016static}.
The analysis can handle dynamic registration, triggering,
and removal of event handlers, but 
not other dynamic nature like dynamic code injection and obfuscation.

\smallskip
\noindent
{\bf Execution Engine}.
Execution engine is used for FSM model construction and event sequence execution. It loads and parses the source file(s), and then executes the top-level
JavaScript code. In particular, for FSM model construction, it interacts with the model constructor by iteratively receiving input events and
outputting an (abstract) successor state and a set of available events at the successor after executing the input event. For event sequence execution, it receives a set of event sequences, executes them one by one and finally outputs a coverage report. For these purpose, we implement an execution engine based on {\sc Artemis}~\cite{artzi2011framework} 
with its features such as event-driven execution model,
interaction with DOM of web pages, dynamic event handlers detection. 

\smallskip
\noindent
{\bf Model Construction}.
The model constructor interacts with the execution engine by iteratively making queries to generate an FSM model up to the given length bound (i.e., Max. Length). The FSM model is intended to represent behaviors of the application. A state of the FSM model denotes an (abstract) state of the application, and a transition
$(s,e,s')$ denotes that after executing the event handler $e$ at the state $s$, the application enters the state $s'$.
The model constructor starts with an FSM containing only one initial state, explores new state $s'$
by selecting one event $e$ available at the current state $s$, adds $(s,e,s')$ into the FSM model,
and continues exploring the state $s'$. \tool allows to restart the exploration from the initial state
and adds new states and transitions into the FSM model, in order to make the FSM model more complete.

\smallskip
\noindent
{\bf Testcase Generation}.
The testcase generator traverses the FSM model to generate event sequences.
\tool supports two event sequence generation algorithms:
(1) partial-order reduction (POR) based event sequence generation, and
(2) random event sequence generation.
The first algorithm traverses the FSM model from the initial state and covers all the paths
up to a (usually small) bound, while redundant event sequences are pruned
based on the POR from {\sc JSDep}~\cite{sung2016static}. It usually generates many short event sequences and is regarded as the baseline algorithm.
The second algorithm repeatedly and randomly traverses the FSM model from the initial state
to generate a small number of longer (up to a usually large bound) event sequences, and, as such,
does not cover all possible paths.

To give a first impression of the performance of \tool, we ran {\sc Artemis}, {\sc JSDep} and \tool on the running example introduced in Section~\ref{runningexample}.
\tool reached 100\% (line) coverage in 0.2s using one event sequence with length $7$, {\sc Artemis} executed 3209 event sequences in 10min and reached 86\%,
and {\sc JSDep} executed 64 event sequences in 0.7s and reached 100\% coverage.
We also ran {\sc JSDep} and \tool on a variant of the running example which contains 10 checkboxes and the condition ${\tt count}>=3$ is replaced by ${\tt count}>=6$.
\tool reached 100\% coverage in 0.4s using one event sequence with length $(10\times 2+1)=21$,
while {\sc JSDep} executed 462 event sequences in 27.4s and reached 100\% coverage.
This suggests that a small number of longer event sequences could outperform a large number of shorter event sequence for applications with intensive DOM event dependency.

\section{Methodology}\label{secl:mothod}
In this section, we present details of our model construction and testcase  generation procedures.
\subsection{Model Construction}\label{sec:modelconst}

\begin{algorithm}[t]
\caption{Model Construction}
\label{alg:mc}
\small
\begin{algorithmic}[1]
\Require
An application $P$ and a Max. bound $d$
\Ensure
An FSM model $M=(S,I, \delta, s_0)$
\State $i:=0$; $I:=\emptyset$; $\delta:=\emptyset$;
\State $cur:={\tt GetInitPage}(P)$;
\State $s_0:={\tt GetState}(cur)$;
\State $S:=\{s_0\}$;
\While{$i <d$}
    \State $e:={\tt GetEvent}(cur)$;
    \State $suc:={\tt GetNextPage}(e)$;
    \State $s':= {\tt GetState}(suc)$;
    \State $\delta:=\delta\cup\{(s_0,e,s')\}; I := I \cup \{e\}$;
    \State $S := S \cup \{s'\}; s_0 := s'$;
    \State $cur:=suc$; $i:=i+1$;
\EndWhile
\State \Return $(S,I, \delta, s_0)$;
\end{algorithmic}
\end{algorithm}

Algorithm~\ref{alg:mc} presents the model construction procedure,
which takes an application $P$ and a maximum bound $d$ as inputs,
and outputs an FSM model $M=(S,I, \delta, s_0)$.
It first calls the function ${\tt GetInitPage}$ which loads and parses $P$, then executes the top-level
JavaScript code, finally returns the initial Web page $cur$ and the set of available events  at this page which are dynamically detected.
The initial state of the FSM model $s_0$ is obtained by calling the function ${\tt GetState}(cur)$.
Intuitively, ${\tt GetState}$ computes a state from the source code of the Web page $cur$ (see below).
After the initialization, Algorithm~\ref{alg:mc} iteratively selects and executes events to explore the state space up to $d$ rounds.
During each iteration, it selects one available event $e$ at the current Web page $cur$ by using
the ${\tt GetEvent}$ function (Line 6), based on some event selection strategy (see below).
Then, it calls ${\tt GetNextPage}$ to get the next Web page $suc$ by executing  $e$. 
The state $s'$ of the new Web page $suc$ is also obtained by calling ${\tt GetState}$. Finally,
a transition $(s,e,s')$ is added into the FSM $M$. Variables $cur$ and $i$ are then updated accordingly. 
As mentioned before,
\tool allows to restart the exploration of the state space from the initial state, so this process may be repeated many times.

Overall, our model construction explores state space in a depth-first fashion with a large length bound (Max. Length in Figure~\ref{Figure:2}),
and can avoid re-executing previously executed event sequences (like {\sc Artemis}~\cite{artzi2011framework}
and {\sc JSDep}~\cite{sung2016static}) or tracking state changes (like {\sc Crawljax}~\cite{Mesbah2012Crawling}),
but not all the paths are explored for efficiency consideration.

\subsubsection{State Abstraction}
As mentioned, states of FSM are used to represent the states of the application and are computed from Web pages via the function ${\tt GetState}$.
State abstraction is crucial to describe application states.
On the one hand, states of FSM should contain enough data to distinguish different application states,
as unexplored application states maybe be wrongly skipped if their abstracted states were explored before, which may happen if a coarse-grained state abstraction is adopted.
On the other hand, over fine-grained state abstraction may generate states which are indistinguishable wrt some coverage criteria, resulting in an explosive or even infinite state-space~\cite{LDD14}.
Therefore, the implementation of ${\tt GetState}$ requires a balance between precision and scalability.

In {\sc Artemis}, the state abstraction is implemented by computing the hash value of the Web page, including Web page layout and dynamically updated DOM tree,
but excluding the concrete values of CSS properties and application variables, and server-side states (unless an initial server state is provided
by the user). It was  demonstrated that this state abstraction is effective and efficient~\cite{artzi2011framework}.


 \begin{figure}[t]
       	\centering
       	\begin{tikzpicture}[->,>=stealth',shorten >=1pt,font=\footnotesize,node distance=1.5cm]       	
       	\node[state,minimum height=0.15cm,minimum width=0.15cm] (s4) {$s_4$};
       	\node[state,minimum height=0.15cm,minimum width=0.15cm] (s2) [above left of=s4] {$s_2$};
       	\node[state,minimum height=0.15cm,minimum width=0.15cm] (s1) [left of=s2] {$s_1$};
       	\node[state, initial,minimum height=0.15cm,minimum width=0.15cm] (s0) [left of=s1] {$s_0$};
       	\node[state,minimum height=0.15cm,minimum width=0.15cm] (s5) [above right of=s4] {$s_5$};
       	\node[state,minimum height=0.15cm,minimum width=0.15cm] (s3) [above of=s4,node distance=2cm] {$s_3$};
       	
       	\path[->]
       		  (s0) edge node [right,above] {A} (s1)
       		  (s1) edge node [right,above] {B} (s2)
       		  (s2) edge node [above left] {C} (s3)
       		  (s2) edge node [below left] {C} (s4)
              (s4) edge node [below right] {Submit} (s5)
        	  (s5) edge node [right,above] {B} (s2)
        	  (s5) edge [loop right] node {A} ();

       	\node[state,initial,minimum height=0.15cm,minimum width=0.15cm] (s6)[below of=s0]   {$s_0$};
       	\path[->]
       	(s6) edge [loop right] node {A,B,C,Submit} () ;
       	
       	\end{tikzpicture}
       	\caption{The FSM models of the running example: the top-part (resp. bottom-part) FSM is constructed using the state abstraction from {\sc Artemis} (resp. our new state abstraction).}
       	\label{fig:FSMmodel}
       \end{figure}
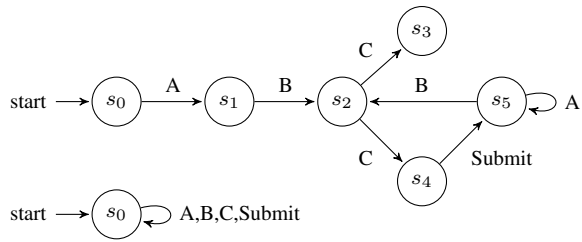
However, our experiments find it is still too fine-grained. {\sc Artemis} typically assigns random values to attributes of DOM nodes
which are taken into account in the state abstraction. 
In our running example, {\sc Artemis} assigns a random Boolean value to the \emph{implicit} attribute {\tt Value} of each checkbox when testing the running example, though
the value of this attribute does not affect the code coverage.
(Here ``implicit" means that the attribute is not explicitly given in source code, but the object has such an attribute.) The FSM model constructed using the state abstraction from {\sc Artemis} is depicted in Figure~\ref{fig:FSMmodel}(top-part). 
To prevent a prohibitively large (sometimes even infinite) state-space and improve efficiency,
we use a state abstraction based on the state abstraction of {\sc Artemis} but discarding the random values of the implicit attributes in the DOM tree. This allows  to focus on the state changes made by executing events  rather than random value assignments. The FSM model constructed by our state abstraction is depicted in Figure~\ref{fig:FSMmodel}(bottom-part), which is much smaller in size.

Experimental results have confirmed that our state abstraction approach significantly reduces the size of FSM models with a comparable line coverage (cf. Section~\ref{sec:s2mc}).

\subsubsection{Event Selection Strategies}
It is common that several events are available at a Web page. 
Evidently the selection of events will affect the quality of the FSM model.
In \tool, we implement two event selection strategies: (1) random event selection as the baseline algorithm and (2) weighted event selection.
The former randomly chooses one of the available events,
namely, ${\tt GetEvent}(cur)$  returns a random available event from the Web page $cur$.
The latter captures the impacts of the previously executed events and DOM event dependencies. Technically we focus on: 
\begin{itemize}
  \item {\bf Frequency of Event Execution.} In principle, all the events should be given opportunities to  execute. As a result, an event which has been executed would have a lower priority to be selected in the subsequent exploration.
We mention that this natural idea was already used to construct an FSM model of Android app for automated testing (e.g. \cite{Su2017stoat}).

 \item {\bf DOM event dependency.} 
    Some corner-case code may be explored only by some specific event sequences due to their dependency. To 
    expose the corner code using long event sequences, the events that \emph{depend upon the previous selected events} deserve a higher chance to be selected.
\end{itemize}
In the weighted event selection strategy, each event $e$ is associated with a weight, which is adjustable dynamically at runtime. The weight of $e$ is defined as follows:
\begin{center}
${\tt weight}(e) = \frac{\alpha_e\times x+ \beta_e\times(1-x)}{N_e+1}$,
\end{center}
where $\alpha_e$ and $\beta_e$ are weight parameters,
$x$ is a Boolean flag determined by DOM event dependency ($x=1$ if $e$ depends upon the previous selected event, $0$ otherwise), $N_e$ is number of times that $e$ has been executed.

With the weighted event selection strategy, ${\tt GetEvent}(cur)$
randomly returns one of events which have the highest weight among all
available events at the Web page $cur$.
In our experiments, $\alpha_e$ and $\beta_e$ are set to be $0.7$ and $0.3$ respectively, which are the best configuration after tuning.

Recall the example in Figure~\ref{Figure:1}. At the initial state,
all the available \emph{onclick} events of {\tt A,B,C} have the same weight $0.3$ (note that the \emph{onclick} event of {\tt Submit} is not available therein), i.e., they have the same chance to be selected.
Suppose the \emph{onclick} event of {\tt A} is selected, the weights of the \emph{onclick} events of {\tt A,B,C} are updated to $\frac{0.7}{2},\frac{0.7}{1},\frac{0.7}{1}$ respectively.
\tool then randomly chooses one of the \emph{onclick} events of {\tt B,C}. Suppose the \emph{onclick} event of {\tt B} is selected on  this occasion.
The weights of all the available \emph{onclick} events of {\tt A,B,C} become $\frac{0.7}{2},\frac{0.7}{2},\frac{0.7}{1}$, which
implies that the \emph{onclick} event of {\tt C} will be selected at the next step. After that,
the weights of the \emph{onclick} events {\tt A,B,C,Submit} are updated to  $\frac{0.7}{2},\frac{0.7}{2},\frac{0.7}{2},\frac{0.7}{1}$ (note that the \emph{onclick} event of {\tt Submit} now becomes available).

\subsection{Testcase Generation}\label{sec:testcasegen}
In this work, as mentioned in the introduction, we focus on the event sequence generation while the
input data is chosen randomly. We first present the baseline algorithm for generating event
sequences with partial-order reduction (POR) which is inspired by~\cite{sung2016static,CYW17}, and then discuss how to generate long test sequences.

\begin{algorithm}[t]
\algsetblock[Name]{If}{EndIf}{}{0.8em}
\algsetblock[Name]{Procedure}{EndProcedure}{}{0.8em}
\algsetblock[Name]{While}{EndWhile}{}{0.8em}
\algsetblock[Name]{For}{EndFor}{}{0.8em}
\algsetblock[Name]{ForAll}{EndFor}{}{0.8em}
\caption{Baseline Event Sequence Generation with POR}
\label{alg:baseline}
\small
\begin{algorithmic}[1]
\Require
\begin{tabular}{l}
           An FSM $M=(S,I, \delta, s_0)$ \\
           A bound $d$ of the length of test sequences \\
           DOM event dependency relation $\rightarrow$
         \end{tabular}
\Ensure
A set of test sequences $T$
\State $T:=\emptyset$; $ss:={\tt New Stack}()$;
\State $ss.{\tt Push}(s_0)$;
\State {\sc Explore}$(ss)$;
\State \Return $T$;
\Procedure{Explore}{${\tt Stack}:ss$}
  \State $s:=ss.{\tt Top}()$;
  \State $s.{\tt SelectedEvent}:={\tt null}$;
  \If{$ss.{\tt Length}() \leq d$}
     \State $s.{\tt done}:=\emptyset$; $s.{\tt sleep}:=\emptyset$;
     \State $E:=\{e\in I\mid \exists s.(e,s')\in \supp(s)\}$;
     \While{$\exists e \in E\setminus(s.{\tt done}\cup s.{\tt sleep})$}
       \State $s.{\tt done}:=s.{\tt done}\cup\{e\}$;
       \State $s.{\tt SelectedEvent}: =e$;
       \ForAll{$s'\in \{s'\in S\mid (e,s')\in \supp(s)\}$}
         \State \textcolor{red}{$s'.{\tt sleep}:=\{e' \in s.{\tt sleep} \mid  e\not\rightarrow e' \wedge  e'\not\rightarrow e \}$};\label{alg:lab:1}
         \State $ss.{\tt push}(s')$;
         \State {\sc Explore}$(ss)$;
         \State \textcolor{red}{$s.{\tt sleep}:= s.{\tt sleep}\cup\{e\}$}; \label{alg:lab:2}
       \EndFor
      \EndWhile
  \EndIf
  \If{$s.{\tt SelectedEvent} = {\tt null}$}
    \State $\rho:=\epsilon$;
    \ForAll{$s'\in ss$ from bottom to top}
      \State $\rho:=\rho\cdot s'.{\tt SelectedEvent}$;
    \EndFor
    \State $T:=T \cup \{\rho\}$;
  \EndIf
  \State $ss.{\tt Pop}()$;
\EndProcedure
\end{algorithmic}
\end{algorithm}

\subsubsection{Baseline Event Sequence Generation with POR}
Given the FSM  $M=(S,I, \delta, s_0)$, a bound $d$, and the DOM event dependency relation $\rightarrow$.
Algorithm~\ref{alg:baseline} (excluding Lines \ref{alg:lab:1} and \ref{alg:lab:2}) generates all possible event sequences with length up to $d$ stored in $T$.

The procedure {\sc Explore} traverses the FSM $M$ in a depth-first manner, where $ss$ is a working stack storing the event sequence with the initial state $s_0$ as the bottom element,
$E$ denotes the set of available events at state $s$, $s.{\tt SelectedEvent}$ denotes the selected event
at $s$, and $s.{\tt done}$ denotes the set of all previously selected events at $s$.
The procedure {\sc Explore} first checks whether the bound $d$ is reached (Line 8),
If $ss.{\tt Length}() > d$, then the while-loop is skipped. After that, if $s.{\tt SelectedEvent} = {\tt null}$ (indicating that the sequence in $ss$ has reached the maximum length $d$), the event sequence
stored in the stack $ss$ is added to the set $T$.
Otherwise, the while-loop will explore a previously unexplored event and invoke the procedure {\sc Explore} recursively.
The while-loop terminates when all the available events at state $s$ have been
explored. Note that in this case, $s.{\tt SelectedEvent} \neq {\tt null}$, hence the sequence in $ss$ will not be added to $T$.

With Lines~\ref{alg:lab:1} and~\ref{alg:lab:2} Algorithm~\ref{alg:baseline}
implements the partial-order reduction based on the notion of sleep-set~\cite{God96}.
In principle, it first classifies
the event sequences into equivalence classes,
and then explores one representative from each equivalence
class. In detail, each event $e$ explored at a state
$s$ is put into its sleep set $s.{\tt sleep}$ (Line 18).
When another event $e'$ is explored at $s$, the sleep set of $s$ is copied into
the sleep set of the next state $s'$ (Line 15), if the DOM events $e$ and $e'$ are independent of each other.
Later, each event at $s$ will be skipped if it is in the sleep set of $s$ (Line 11), because executing this event is guaranteed
to reach a previously explored state.

\subsubsection{Long Event Sequence Generation}

Algorithm~\ref{alg:randomltc} shows the pseudocode of our random event sequence generation.
Given the FSM model $M=(S,I, \delta, s_0)$, a maximum length bound $d$ and a maximum bound $m$ of the number of event sequences,
Algorithm~\ref{alg:randomltc} randomly generates $m$ number of event sequences with length $d$.
Each iteration of the outer while-loop computes one event sequence with length $d$ until the number of event sequences reaches $m$.
In the inner while-loop, it starts from the initial state $s_0$ and an empty sequence $\rho=\epsilon$.
At each state $s$,
the inner while-loop iteratively and randomly selects a pair  $(e,s')$ of event and state denoting that executing the event $e$
at the state $s$ moves to the state $s'$, then appends $e$ to the end of previously computed sequence $\rho$.
Algorithm~\ref{alg:randomltc} repeats this procedure until the length of $\rho$ reaches the maximum bound $d$.
At this moment, one event sequence is generated and stored into the set $T$.
The outer while-loop enters its next iteration.

Algorithm~\ref{alg:randomltc} may not generate all the possible event sequences, and may generate redundant event sequences, but less often, due to large maximum bound.
We remark that the POR technique cannot be integrated into Algorithm~\ref{alg:randomltc}.

\begin{algorithm}[t]
\caption{Long Event Sequence Generation}
\label{alg:randomltc}
\small
\begin{algorithmic}[1]
\Require
\begin{tabular}{l}
           An FSM $M=(S,I, \delta, s_0)$ \\
           A bound $d$ of the length of test sequences  \\
           A bound $m$ of the number of test sequences
         \end{tabular}
\Ensure
A set of test sequences $T$
\State $T:=\emptyset$;
\While{$|T| < m$}
    \State $\rho:=\epsilon$;
    \State $s:=s_0$;
    \While{$|\rho| < d$}
        \State $(e,s):={\tt RandomlySelectOnePair}(\supp(s))$;
        \State $\rho:=\rho\cdot e$;
    \EndWhile
    \State $T:=T\cup \{\rho\}$;
\EndWhile
\State \Return $T$;
\end{algorithmic}
\end{algorithm}

\section{Experiments}\label{sec:exp}
\label{sec:eval}
We have implemented our method as a software tool  \tool.
It exploits {\sc JSDep}~\cite{sung2016static} for computing DOM event dependency and
a modified automated testing framework {\sc Artemis}~\cite{artzi2011framework} as the execution engine. 
For comparison purpose, we implemented \tool
in such a way that individual techniques are modularized and can be
enabled on demand. 
Thus, we were able to compare the performance of various approaches with different configurations, in particular, (1) our state abstraction vs the state abstraction from~\cite{artzi2011framework},
(2) random event selection vs weighted event selection,
(3) baseline event sequence generation with POR (i.e., Algorithm~\ref{alg:baseline}) vs long event sequence generation (i.e., Algorithm~\ref{alg:randomltc}).
To demonstrate the efficiency and effectiveness of  \tool, we compared \tool with {\sc JSDep}~\cite{sung2016static} on same
benchmarks. (We note that in~\cite{sung2016static} {\sc JSDep} is shown to be superior to {\sc Artemis}~\cite{artzi2011framework}, so a direct comparison between \tool and {\sc Artemis}~\cite{artzi2011framework}
is excluded.)

The experiments are designed to answer the following research questions:

\begin{enumerate}[label=\textbf{RQ\arabic*.},itemindent=*,leftmargin=*]
    \item How efficient and effective is \tool compared with {\sc JSDep}~\cite{sung2016static}?
    \item How effective is our coarse-grained state abstraction compared with the state abstraction from~\cite{artzi2011framework}?
    \item How effective is the weighted event selection strategy compared with the random event selection strategy?
	\item How effective is the long event sequence generation compared with the baseline algorithm?
\end{enumerate}

\subsection{Evaluation Setup}
To make comparison on a fair basis,
we evaluated \tool on publicly available benchmarks\footnote{\url{https://github.com/ChunghaSung/JSdep}.} of {\sc JSDep}~\cite{sung2016static}, which consist of 21 client-side JavaScript Web applications with 18,559 lines of code in total.
Columns 1-2 of Table~\ref{tab:comOfLongJS} show the name of the application and the number of lines of code.
We ran all experiments on a server with a 64-bit Ubuntu 12.04 OS, Intel Xeon(R) E5-2603v4 CPU (1.70 GHz, 6 Cores), and 32GB RAM. To answer the research questions {\bf RQ1}-{\bf RQ4}, we conducted four case studies.
The time used to compute the DOM event dependency  is usually marginal and can be safely ignored, so is not counted in line with \cite{sung2016static}. For statistics, we ran \tool on each application 5 times and simply took the average as the result.
The coverage measure are the aggregation of that from model construction and testcase execution respectively which are separated in the last experiment.

\subsection{{\bf RQ1:} \tool vs. {\sc JSDep}}
\begin{table}
\centering
\setlength{\tabcolsep}{2.2pt}
\begin{tabular}{|l|r|rrrl|rrr|}
\hline
\multirow{2}{*}{Name} & \multirow{2}{*}{Loc} & \multicolumn{4}{c|}{\tool}                           & \multicolumn{3}{c|}{JSDep}                     \\
\cline{3-9}
                      &                      & CRG.      & sd        & Tests        & Len.    & CRG.            & Tests        & M.Len.        \\
\hline
case1                 & 59                   & 100.0\%   & 0          & 4679        & 99       & 100.0\%         & 1409         & 705           \\
case2                 & 72                   & 100.0\%   & 0          & 4376        & 99      & 100.0\%         & 3058         & 549           \\
case3                 & 165                  & 100.0\%    & 0        & 2739        & 99        & 100.0\%         & 7811         & 575           \\
case4                 & 196                  & \textbf{87.0\%}  & 0  & 2816        & 99       & \textbf{77.9\%}  & 8594         & \textbf{500}  \\
\textbf{frog}         & 567                  & \textbf{96.8\%}  & 0.004  & \textbf{15} & 99   & \textbf{84.6\%} & \textbf{86}  & \textbf{16}   \\
cosmos                & 363                  & 82.0\%           & 0.060  & 322         & 99  & 79.5\%          & 973          & 243           \\
hanoi                 & 246                  & 89.0\%           & 0   & 1303        & 99       & 82.5\%          & 902          & 225           \\
flipflop              & 525                  & 97.0\%           & 0   & 59          & 99       & 96.3\%          & 284          & 71            \\
\textbf{sokoban}      & \textbf{3056}        & \textbf{88.6\%}  & 0.026  & \textbf{58} & 99   & \textbf{77.6\%} & \textbf{203} & \textbf{51}   \\
wormy                 & 570                  & 45.4\%           & 0.054    & 35          & 99   & 41.0\%          & 323          & 18            \\
chinabox              & 338                  & 84.0\%           & 0   & 7           & 99        & 82.3\%          & 92           & 9             \\
\textbf{3dmodel}      & \textbf{5414}        & \textbf{85.0\%}  & 0   & \textbf{8}  & 99        & \textbf{71.5\%} & \textbf{66}  & \textbf{10}   \\
\textbf{cubuild}      & \textbf{1014}        & \textbf{88.8\%}  & 0.060& \textbf{7}  & 99    & \textbf{72.8\%} & \textbf{153} & \textbf{17}   \\
pearlski              & 960                  & 55.0\%           & 0   & 72          & 99       & 54.9\%          & 214          & 52            \\
speedyeater           & 784                  & 89.8\%            & 0.010 & 516         & 99   & 82.1\%          & 1497         & 374           \\
gallony               & 300                  & 95.0\%           & 0 & 2133        & 99         & 94.5\%          & 1611         & 95            \\
fullhouse             & 528                  & 92.2\%           & 0.012 & 1119        & 99   & 86.3\%          & 889          & 222           \\
\textbf{ball\_ool}    & \textbf{1745}        & \textbf{92.8\%}  & 0.004  & \textbf{1}  & 99   & \textbf{74.2\%} & \textbf{18}  & \textbf{4}    \\
harehound             & 468                  & 94.8\%          & 0.004   & 522         & 99   & 94.5\%          & 1224         & 116           \\
match                 & 369                  & 72.8\%          & 0.004   & 1063        & 99    & 73.2\%          & 4050         & 845           \\
lady                  & 820                  & 79.0\%           & 0    & 0           & 99      & 75.7\%          & 35           & 8             \\
\hline
\textbf{Average}      & 883.8                & \textbf{86.4\%}  & 0.011 & 1040.5      & 99   & \textbf{81.0\%} & 1594.9       & -             \\
\hline
\end{tabular}
\caption{Coverage of \tool and {\sc JSDep} in 600s.\label{tab:comOfLongJS}}
\vspace{-2em}
\end{table}

\begin{figure*}
	\centering
	\subfloat{\includegraphics[width=0.33\textwidth]{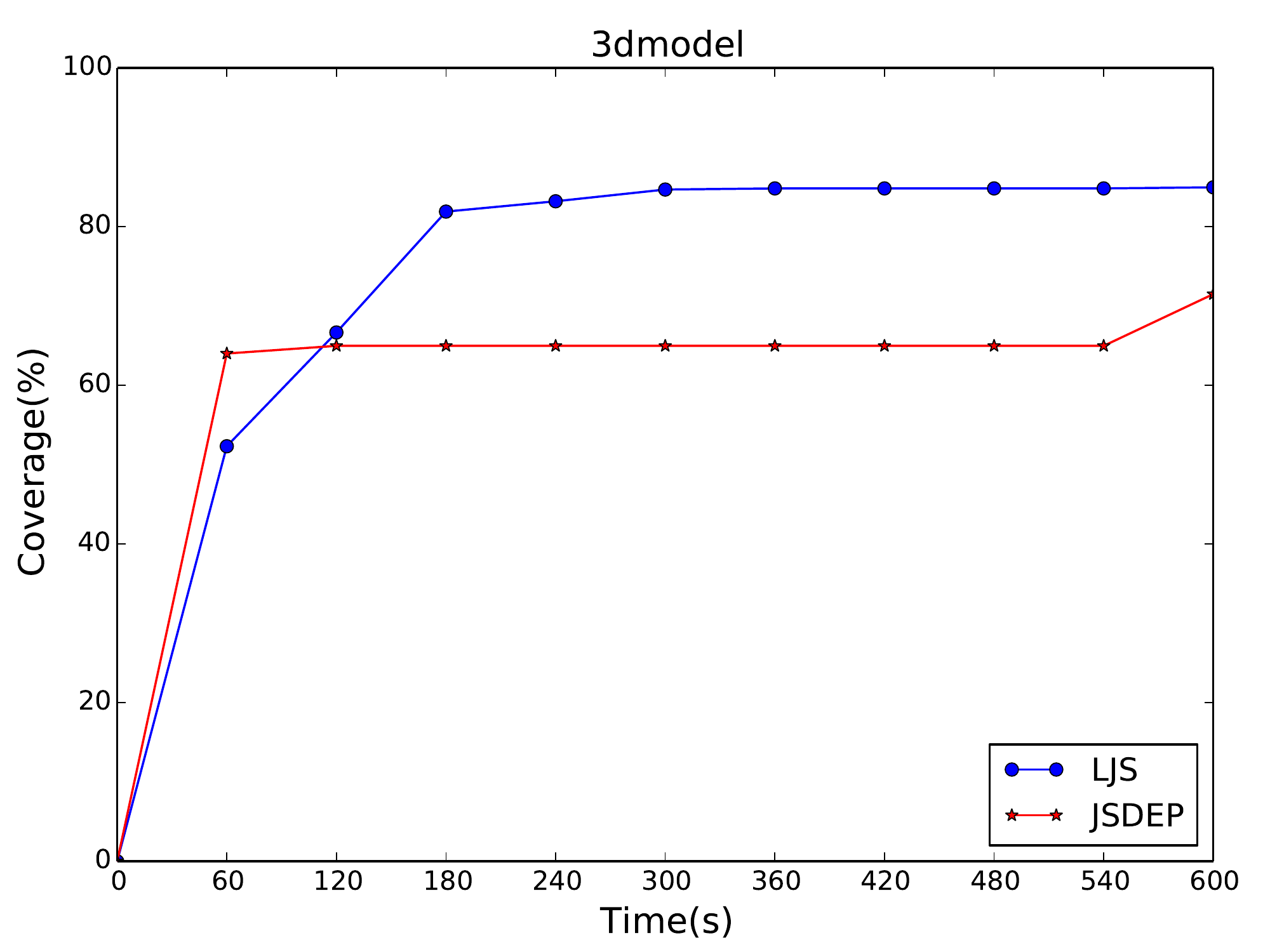}}
	\subfloat{\includegraphics[width=0.33\textwidth]{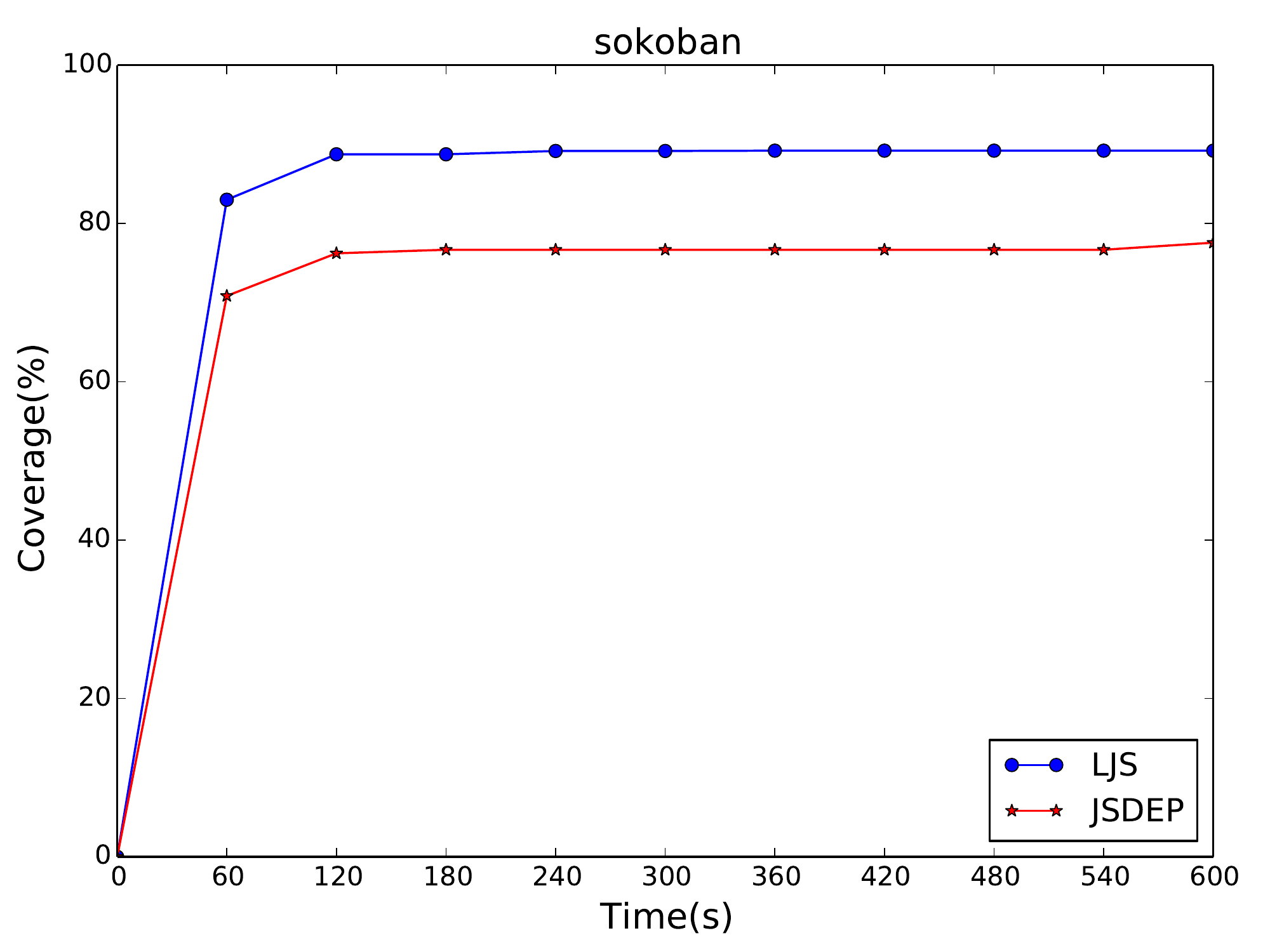}}
	\subfloat{\includegraphics[width=0.33\textwidth]{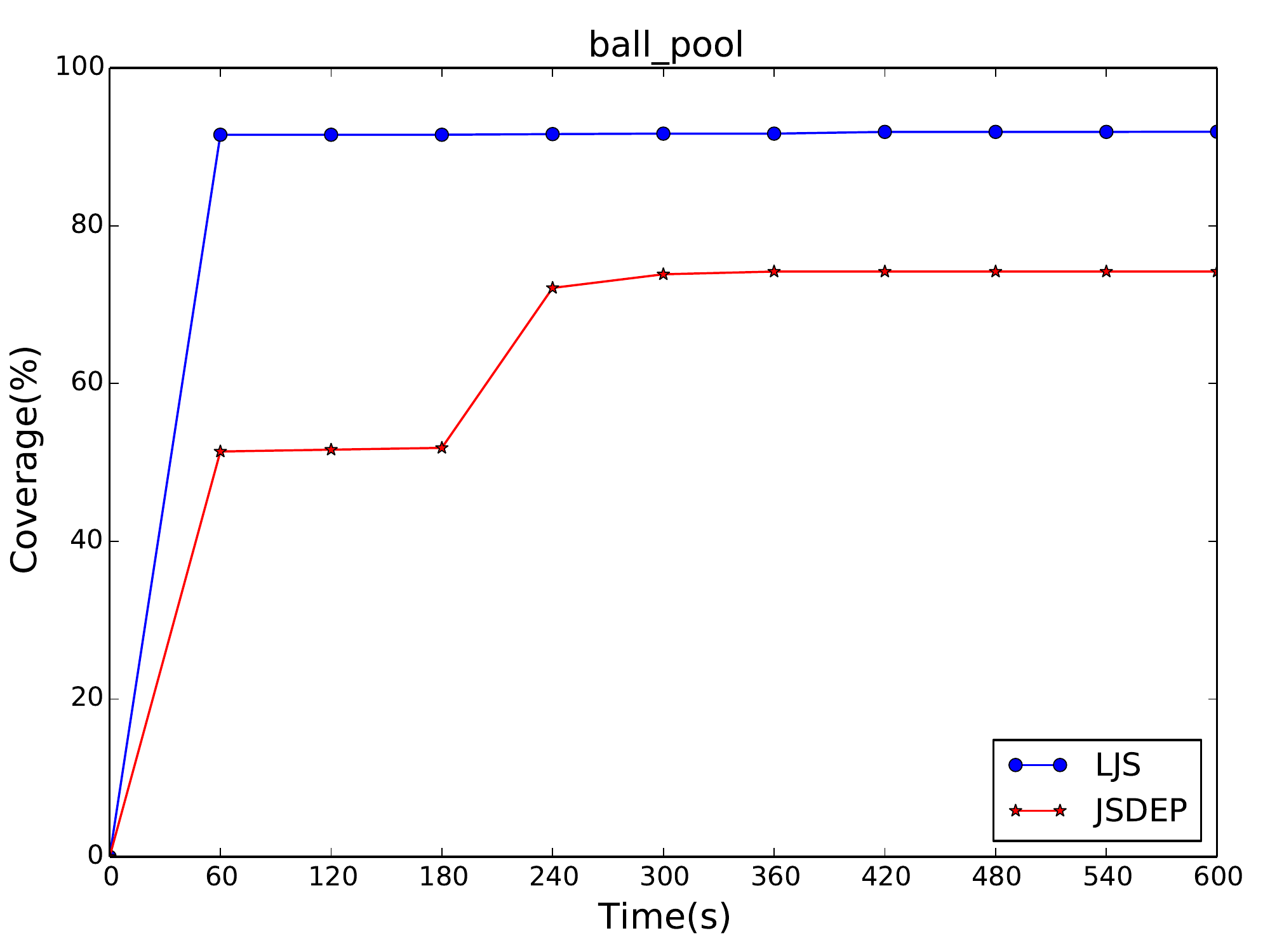}}\
	\subfloat{\includegraphics[width=0.33\textwidth]{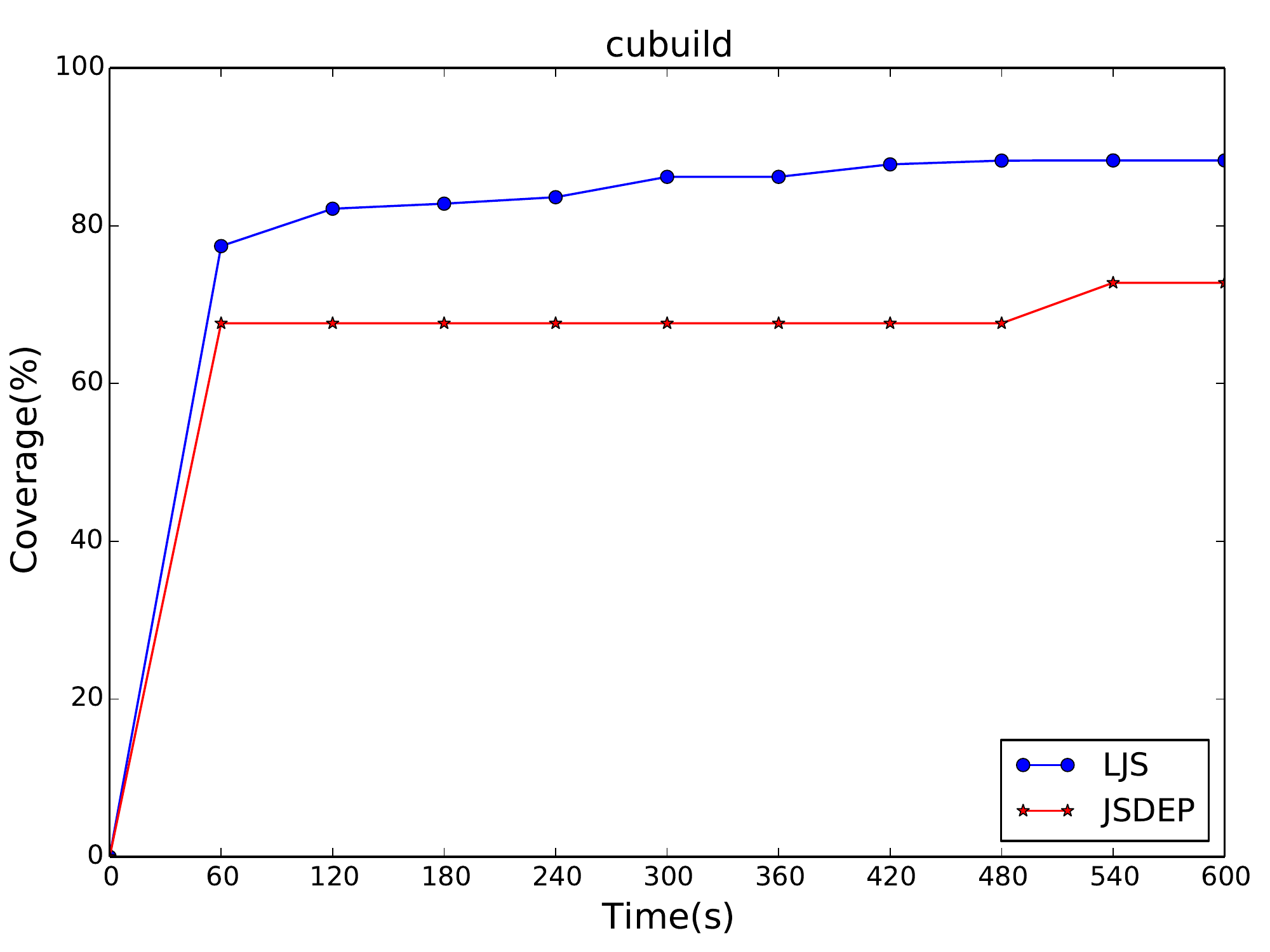}}
	\subfloat{\includegraphics[width=0.33\textwidth]{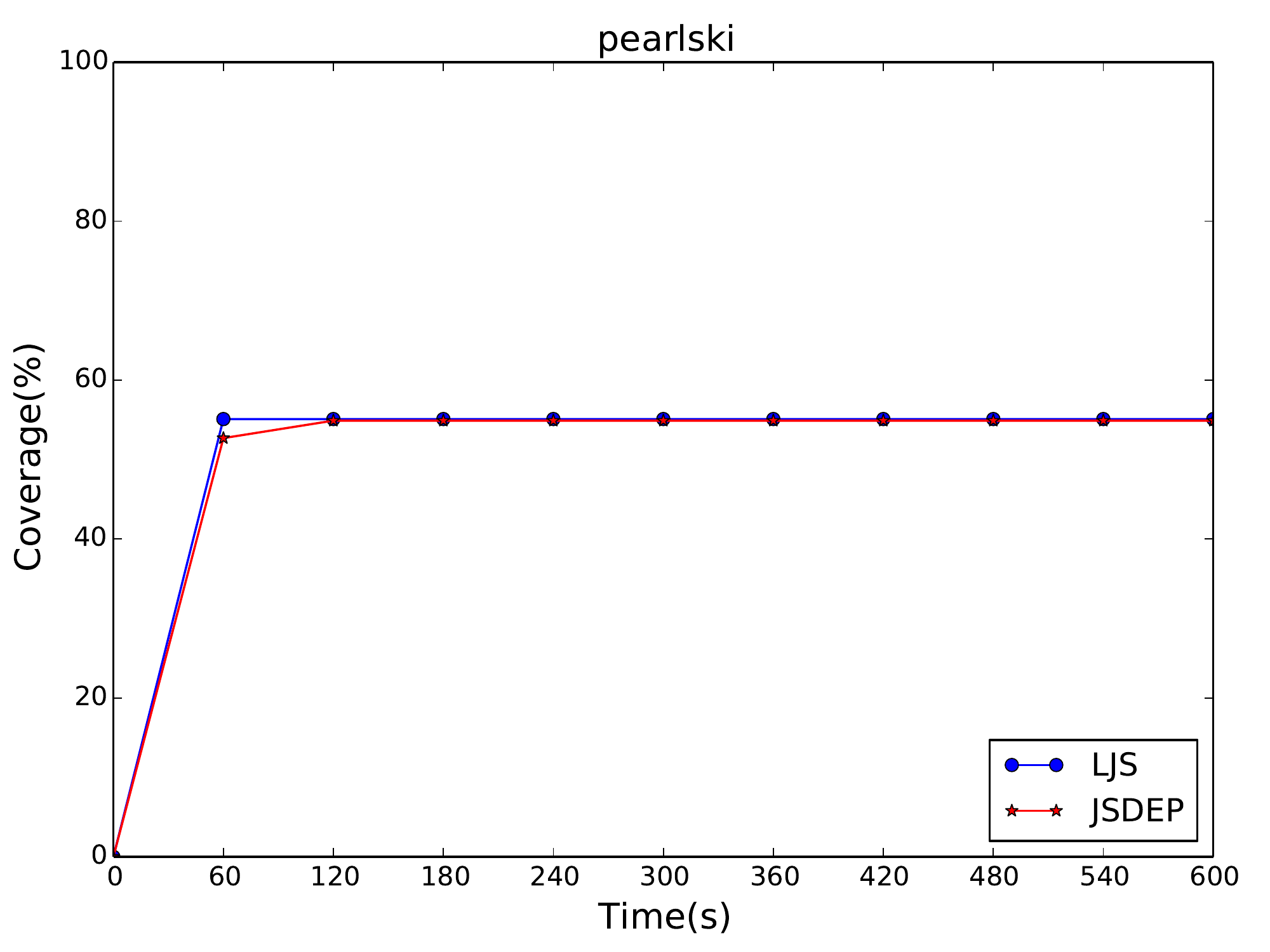}}
	\subfloat{\includegraphics[width=0.33\textwidth]{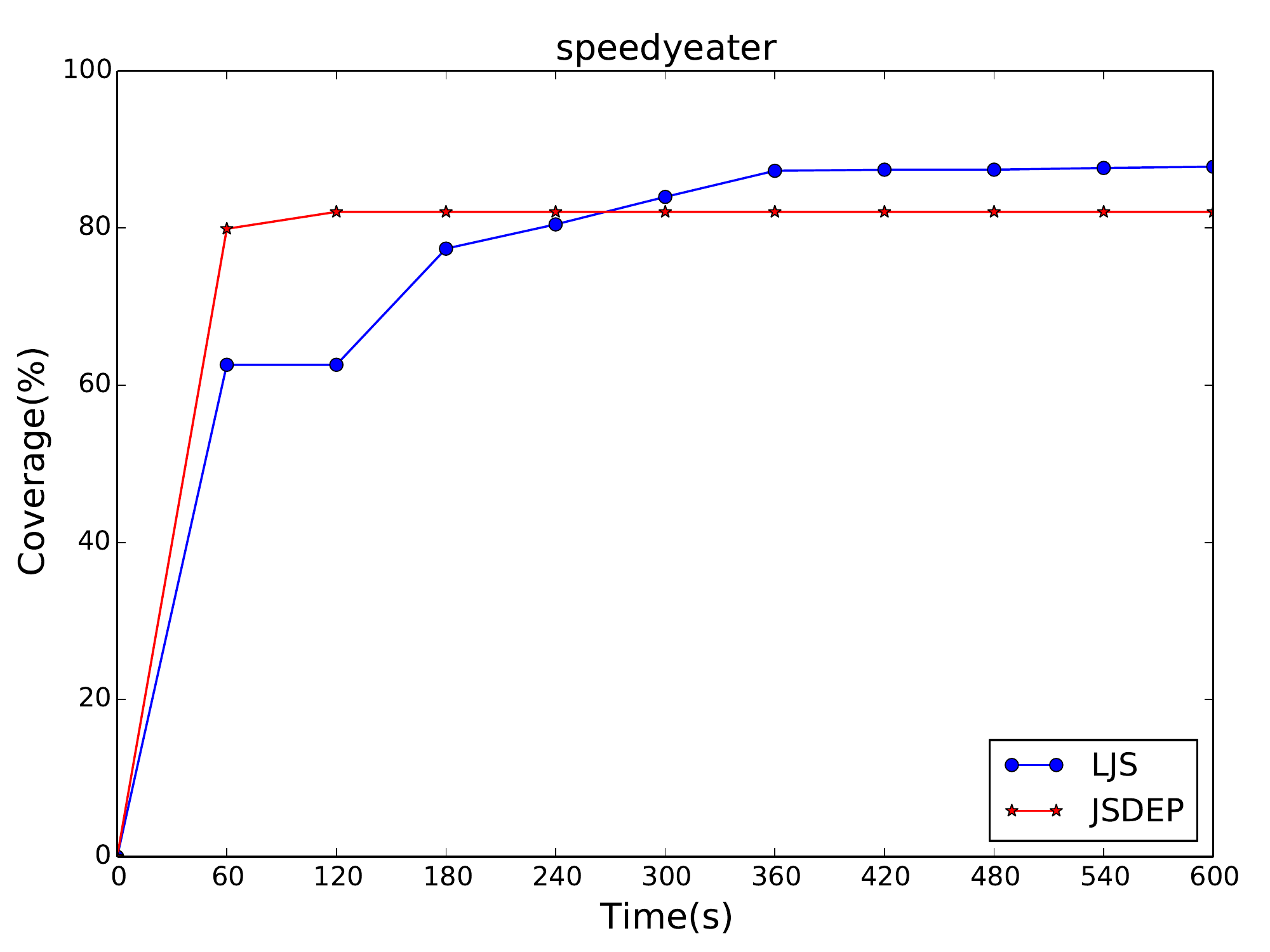}}
     \caption{Coverage of \tool and {\sc JSDep}, as a function of the execution time.}
	\label{fig:com}
\end{figure*}

In this study, we answer {\bf RQ1} by performing two experiments assessing effectiveness and efficiency.
For effectiveness, we compare the (line) coverage of \tool with {\sc JSDep} in 600 seconds.
For efficiency, we compare the time used by \tool and {\sc JSDep} to achieve the same coverage.
The experiments of \tool were performed with the following configurations: Max. Length=99, 
random event selection strategy, our new state abstraction, running Algorithm~\ref{alg:mc} two times, and long event sequence generation.
Experiments of {\sc JSDep} were performed with the setting shown in~\cite{sung2016static}.

Table~\ref{tab:comOfLongJS} shows the results of \tool and {\sc JSDep} in 600s,
where 
Columns 3-6 (resp.\ Columns 7-9) show the average, standard deviation (sd) of coverage obtained in 5 runs, number and length of event sequences after running \tool (resp. {\sc JSDep}). 

Overall, we can observe an increase in the average coverage from 81\% achieved by {\sc JSDep} to 86.4\%
achieved by \tool. (We remark that the average coverage was increased from 67\% achieved by {\sc Artemis} to 80\% achieved by {\sc JSDep} in 600s~\cite{sung2016static}. In our experiment, {\sc JSDep} performed slightly better.)
Perhaps more importantly, on large applications such as \emph{frog}, \emph{sokoban}, \emph{3dmodel}, \emph{cubuild} and \emph{ball\_pool}, the coverage of \tool is 11-18\% higher than that of {\sc JSDep}. We can also observe that a small amount of long event sequences could outperform a large amount of short event sequences.
However, it should be emphasized that long event sequences, but of low quality, may not improve the coverage.
This has been demonstrated by the results of \emph{case4}, \emph{speedyeater} and \emph{fullhouse}.


Figure~\ref{fig:com} shows the coverage that \tool and {\sc JSDep} achieved by running on the top 6 largest applications in time ranging from 60s to 600s with step size 60s. (Note that the application \emph{lady} is excluded in this experiments because its  model construction takes more than 600s; cf. Table~\ref{tab:comOfLongJS}.) The X-axis is the execution time budget whereas Y-axis is the achieved coverage.
Overall, as the execution time budget increases, the coverage of \tool is higher than that of {\sc JSDep}. Moreover, the rate of \tool to achieve a higher coverage
is, in most cases, slightly higher than that of {\sc JSDep}.

\subsection{{\bf RQ2:} Comparison of State Abstraction Techniques}
\label{sec:s2mc}
\begin{table}[t]
	\centering
\setlength{\tabcolsep}{1.8pt}
	\begin{tabular}{|l|rrrr|rrrr|}
		\hline
		\multirow{2}{*}{Name} & \multicolumn{4}{c|}{State abstraction from~\cite{artzi2011framework}}    & \multicolumn{4}{c|}{Our state abstraction}     \\ \cline{2-9}
		& CRG. & $|S|$ & $|\delta|$ & Time(s)  & CRG. &  $|S|$ & $|\delta|$ & Time(s)   \\ \hline
		case1           & 100.0\%  & 1.0           & 2.0          & 0.5          & 100.0\%     & 1.0           & 2.0          & 0.0                    \\
		case2           & 97.6\%   & 1.0           & 4.0          & 0.7          & 95.2\%      & 1.0           & 4.0          & 0.9                 \\
		case3           & 100.0\%  & 1.0           & 6.0          & 1.2          & 99.0\%      & 1.0           & 6.0          & 1.4                    \\
		case4           & 87.0\%  & 1.0           & 8.0          & 1.3          & 87.0\%     & 1.0           & 8.0          & 1.5                    \\
		{\bf frog}      & 95.2\%   & {\bf 199.0}   & {\bf 198.0}  & 87.5         & 94.8\%      & {\bf 24.2}    & {\bf 100.8}  & 105.9               \\
		{\bf cosmos}    & 79.0\%   & {\bf 125.0}   & {\bf 196.0}  & 9.0          & 78.8\%      & {\bf 65.6}    & {\bf 142.2}  & 10.1                \\
		hanoi           & 89.0\%   & 104.8         & 186.0        & 2.0          & 89.0\%      & 99.0          & 185.4        & 2.3                    \\
		flipflop        & 97.4\%   & 25.8          & 100.4        & 21.6         & 97.0\%      & 27.2          & 110.8        & 18.5                   \\
		{\bf sokoban}   & 88.4\%   & {\bf 68.6}    & {\bf 191.4}  & 23.6         & 88.6\%      & {\bf 31.2}    & {\bf 125.8}  & 26.4                \\
		wormy           & 42.2\%   & 189.8         & 198.0        & 37.9         & 40.8\%      & 132.8         & 185.4        & 38.6                \\
		chinabox        & 84.0\%   & 67.6          & 156.4        & 136.8        & 84.0\%      & 69.8          & 161.4        & 151.2                  \\
{\bf 3dmodel}         & 81.6\%   & {\bf 3.0}      & {\bf 25.6}         & 105.5        & 83.2\%      & {\bf 1.0} & {\bf 6.0}     & 106.2               \\
		cubuild         & 85.0\%   & 87.0          & 166.6        & 131.8        & 84.8\%      & 87.4          & 163.2        & 132.5               \\
{\bf pearlski}      & 55.0\%   & {\bf 136.8}   & {\bf 196.4}  & 18.2         & 55.0\%      & {\bf 75.6}    & {\bf 176.8}  & 19.2                   \\
{\bf speedyeater}   & 82.0\%   & {\bf 175.8}   & {\bf 198.0}  & 6.0          & 81.2\%      & {\bf 4.6}     & {\bf 65.0}   & 9.5                 \\
		gallony         & 95.0\%   & 63.8          & 174.8        & 1.4          & 95.0\%      & 63.4          & 176.0        & 1.7                    \\
{\bf fullhouse}     & 93.0\%   & {\bf 161.6}   & {\bf 198.0}  & 2.5          & 93.0\%      & {\bf 28.6}    & {\bf 116.0}  & 3.3                    \\
ball\_pool          & 93.0\%   & 40.6          & 142.2        & 354.4        & 93.0\%      & 39.8          & 150.4        & 346.2                  \\
{\bf harehound}     & 83.8\%   & {\bf 191.0}   & {\bf 198.0}  & 5.9          & 84.8\%      & {\bf 16.2}    & {\bf 109.4}  & 9.0                 \\
{\bf match}         & 67.4\%   & {\bf 13.6}    & {\bf 171.0}  & 3.2          & 69.8\%      & {\bf 4.4}     & {\bf 62.8}   & 5.0                 \\
{\bf lady}          & 79.2\%   & {\bf 173.4}   & {\bf 197.8}  & 936.1        & 79.2\%      & {\bf 85.6}    & {\bf 174.0}  & 1109.6              \\ \hline
{\bf Average}       &84.5\%    &{\bf 87.2}     &{\bf 138.8}   & 89.9         &84.4\%       &{\bf 41.0}     &{\bf 106.3}   & 100.0           \\ \hline
	\end{tabular}
\caption{Comparison of state abstraction techniques.\label{Table:aon}}
\vspace{-2em}
\end{table}

In this study, we answer {\bf RQ2} by performing one experiment and comparing the obtained FSM model and coverage results using our coarse-grained  state abstraction and the state abstraction from~\cite{artzi2011framework} respectively. In this experiment, \tool constructs the FSM model by running Algorithm~\ref{alg:mc} two times using Max. Length=99 
and random event selection, and generates two event sequences from
the FSM model. Taking into account the consumption of time, the experiment only generates two event sequences.

Table~\ref{Table:aon} shows the results, where 
Columns 2-5 (resp. Columns 6-9) show the coverage, numbers of states and transitions of the FSM model (note that we take the average of these numbers), and execution time after running \tool
with the state abstraction from~\cite{artzi2011framework} (resp. our new state abstraction).


Overall,  the numbers of states and transitions using our state abstraction are much smaller, with a dramatic
decrease in some large applications such as \emph{sokoban}, \emph{3dmodel}, \emph{pearlski}, \emph{speedyeater} and \emph{harehound}. 
Meanwhile, the performance of the two state abstractions (in terms of average coverage and execution time) is
comparable.

\subsection{{\bf RQ3:} Comparison of Event Selection Strategies}

\begin{table}[t]
	\centering
\setlength{\tabcolsep}{1.8pt}
		\begin{tabular}{|l|rrrr|rrrr|}
			\hline
\multirow{2}{*}{Name} & \multicolumn{4}{c|}{Random Event Selection}    & \multicolumn{4}{c|}{Weighted Event Selection} \\ \cline{2-9}
			& CRG. & $|S|$ & $|\delta|$ & Time(s) & CRG. & $|S|$ & $|\delta|$ & Time(s)  \\
			\hline
case1       & 100.0\%     & 1.0    & 2.0      & 0.5     & 100.0\%    & 1.0   & 2.0     & 0.6  \\
case2       & 100.0\%     & 1.0    & 4.0      & 0.9     & 97.6\%   & 1.0   & 4.0     & 1.4  \\
case3       & 98.0\%      & 1.0    & 6.0      & 1.9     & 100.0\%    & 1.0   & 6.0     & 2.2  \\
case4       & 87.0\%     & 1.0    & 8.0      & 1.4     & 87.0\%    & 1.0   & 8.0     & 1.7  \\
frog        & 95.2\%      & 28.6   & 101.0    & 107.1   & 95.2\%   & 52.0  & 119.8   & 208.9\\
cosmos      & 79.0\%      & 66.6   & 143.6    & 9.5     & 76.8\%   & 24.0  & 149.0   & 9.0  \\
hanoi       & 89.0\%      & 102.8  & 186.4    & 2.1     & 89.0\%     & 112.0 & 121.4   & 3.5  \\
flipflop    & 97.0\%      & 26.6   & 108.2    & 20.3    & 97.0\%     & 17.0   &55.8     &50.6 \\
sokoban     & 89.6\%      & 28.6   & 123.2    & 24.8    & 84.0\%     & 9.0   & 36.0    & 25.4 \\
wormy       & 40.6\%      & 131.8  & 186.8    & 37.1    & 41.8\%   & 134.0 & 174.0   & 37.8 \\
chinabox    & 84.0\%      & 60.0   & 151.6    & 159.2   & 84.0\%     & 72.2  & 159.8   & 129.4\\
{\bf 3dmodel} & {\bf 82.0\%}      & 1.0    & 6.0      & {\bf 98.1}    & {\bf 84.0\%}     & 1.0   & 6.0     & {\bf 89.1} \\
{\bf cubuild} & {\bf 84.8\%}  & {\bf 92.8}   & {\bf 171.2}  & {\bf 136.9}   & {\bf 85.6\%}  & {\bf 85.4}  & {\bf 141.4} & {\bf 130.1}\\
pearlski    & 55.0\%      & 71.2   & 175.0    & 18.9    & 51.0\%     & 33.0  & 87.0    & 17.5 \\
speedyeater & 87.6\%      & 4.8    & 65.2     & 8.9     & 82.2\%   & 2.0   & 34.0    & 9.9  \\
gallony     & 95.0\%      & 61.6   & 173.8    & 1.5     & 92.0\%     & 30.4  & 127.6   & 2.1  \\
{\bf fullhouse}   & {\bf 93.0\%}      & 30.6   & 123.0    & 3.1     & {\bf 75.0\%}     & 5.0   & 13.0    & 4.7  \\
ball\_pool  & 93.4\%      & 42.2   & 146.2    & 345.1   & 92.8\%   & 54.6  & 191.0   & 404.9\\
harehound   & 89.4\%      & 11.8   & 99.2     & 9.1     & 88.2\%   & 3.0   & 34.0    & 10.4 \\
match       & 69.4\%      & 3.6    & 56.4     & 4.8     & 69.6\%   & 4.0   & 50.6    & 5.6 \\
lady        &   79.0\%     & 82.6   & 171.4   &  1849.2         & 78.0\%   & 67.4  & 153.6 & 1881.1 \\ \hline
       {\bf Average}  &85.1\%  &{\bf 40.5}  &{\bf 105.2}  &135.3  &83.4\%  &{\bf 33.8}  &{\bf 79.7}  &144.1 \\ \hline
	\end{tabular}
	\caption{Comparison of event selection strategies.\label{Table:wor}}
\vspace{-2em}
\end{table}

In this study, we answer {\bf RQ3} by performing one experiment and comparing the obtained FSM model and coverage results using weighted event selection strategy and random event selection strategy respectively.
In this experiment, \tool constructs the FSM model by running Algorithm~\ref{alg:mc} two times using 
Max. Length=99 and our coarse-grained state abstraction, and generates two event sequences from
the FSM model.

Table~\ref{Table:wor} shows the results, where 
Columns 2-5 (resp. Columns 6-9) show the coverage, numbers of states and transitions of the FSM model, and execution time after running \tool with random event selection (resp. weighted event selection).
%

Overall, the numbers of states and transitions of weighted event selection are respectively less than that of random event selection. Meanwhile, the average coverage and  execution time of the two selection strategies are
comparable. In particular, in terms of coverage for applications such as \emph{cubuild} the weighted event selection strategy performs better, whereas for applications such as \emph{fullhouse}, the random event selection strategy performs better. We will discuss these findings later.

\subsection{{\bf RQ4:} Comparison of Event Sequence Generations}

\begin{figure}
	\includegraphics[width=0.5\textwidth]{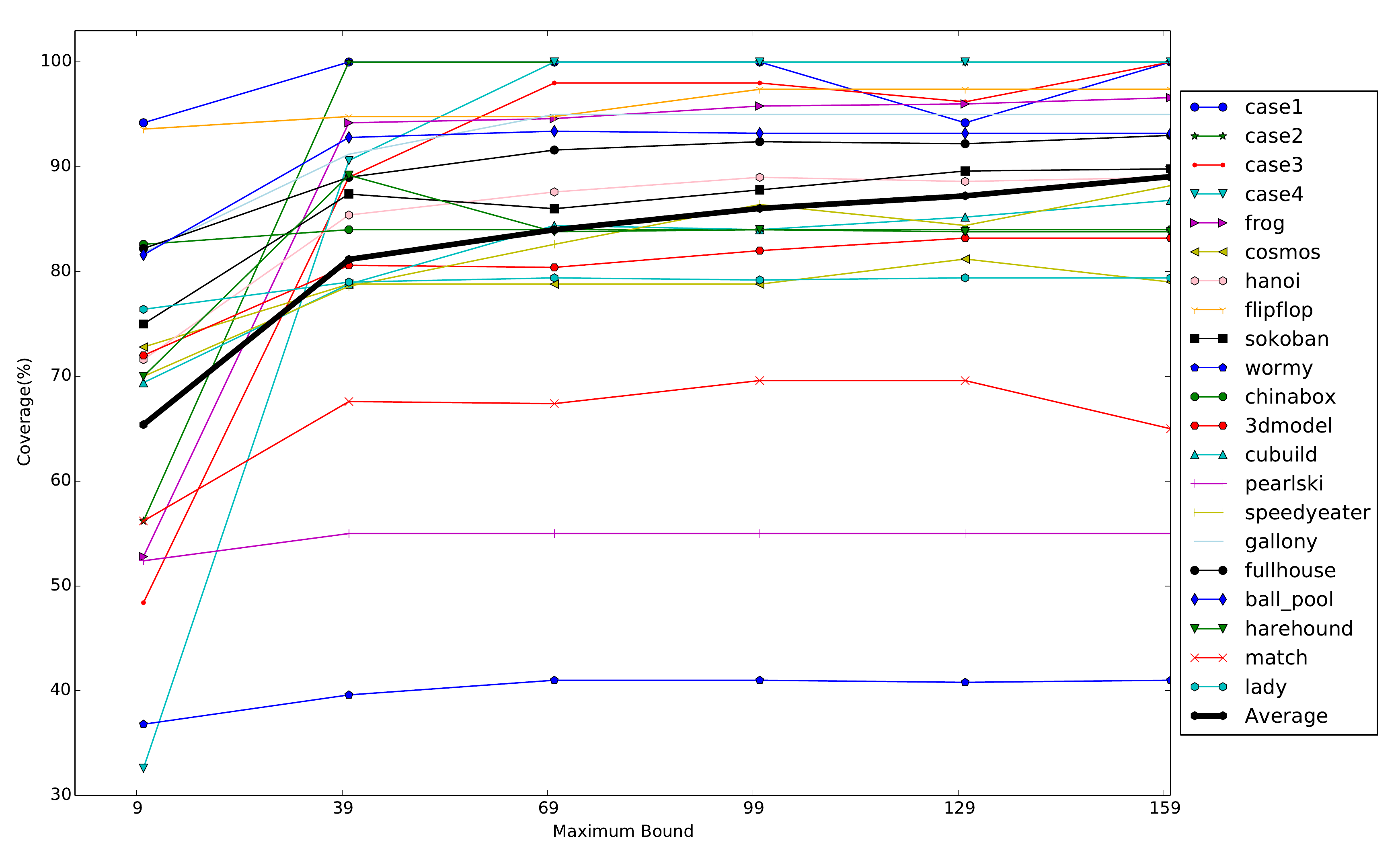}
	\caption{Coverage of \tool, as a function of the length bound.\label{Figure:sd}}
\end{figure}

\begin{table}[t]
\setlength{\tabcolsep}{2pt}
	\centering
		\begin{tabular}{|l|rrrr|rrrr|}
\hline
\multirow{2}{*}{Name} & \multicolumn{4}{c|}{Baseline with POR}      & \multicolumn{4}{c|}{Long Sequence Generation}                                                                                                 \\
\cline{2-2}\cline{3-3}\cline{4-4}\cline{5-9}
& CRG. & Len. & Time(s) & Tests  & CRG.& Len. & Times & Tests  \\
			\hline
case1            & 100.0\%        & 16  & 829          & 65504          & 100.0\%       & 99     & 600     & 4543           \\
{\bf case2}      & {\bf 87.8\%}   & 9   & 1125         & {\bf 87040}    & {\bf 100.0\%} & 99     & 600     & 4244           \\
{\bf case3}      & {\bf 68.6\%}   & 7   & 931          & {\bf 54432}    & {\bf 100.0\%} & 99     & 600     & 2658           \\
{\bf case4}      & {\bf 61.8\%}   & 6   & 674          & {\bf 32768}    & {\bf 87.0\%} & 99     & 600     & 2732           \\
frog             & 88.6\%         & 4   & 664          & 592            & 95.8\%        & 99     & 600     & 8              \\
cosmos           & 78.0\%         & 5   & 1072         & 4415           & 83.8\%        & 99     & 600     & 175            \\
hanoi            & 86.8\%         & 6   & 889          & 4276           & 89.0\%        & 99     & 600     & 1365           \\
flipflop        & 97.0\%          & 6   & 1200         & 419            & 97.0\%        & 99     & 600     & 66             \\
sokoban         & 88.6\%          & 5   & 1200         & 1697           & 89.0\%        & 99     & 600     & 54             \\
wormy           & 41.2\%          & 13  & 657          & 317            & 43.2\%        & 99     & 600     & 30             \\
{\bf chinabox}   & {\bf 78.0\%}   & 8   & {\bf 729}    & {\bf 64}       & {\bf 84.0\%}  & 99     & 600     & {\bf 5}          \\
{\bf 3dmodel}   & {\bf 72.0\%}    & 4   & {\bf 1200}   & {\bf 93}       & {\bf 85.0\%}  & 99     & 600     & {\bf 10}         \\
{\bf cubuild}    & {\bf 75.2\%}   & 6   & {\bf 1019}   & {\bf 238}      & {\bf 92.2\%}  & 99     & 600     & {\bf 10}         \\
pearlski         & 52.1\%         & 7   & 960          & 653            & 55.0\%        & 99     & 600     & 85             \\
speedyeater       & 82.3\%        & 5   & 1200         & 17106          & 88.0\%        & 99     & 600     & 508            \\
gallony         & 94.5\%          & 8   & 1200         & 9821           & 95.0\%        & 99     & 600     & 2852           \\
fullhouse        & 79.2\%         & 8   & 1200         & 12097          & 79.0\%        & 99     & 600     & 1261           \\
ball\_pool       & 92.1\%         & 6   & 1200         & 22             & 93.0\%        & 99     & 600     & 3              \\
harehound       & 92.2\%          & 4   & 1200         & 7949           & 95.0\%        & 99     & 600     & 420            \\
match            & 73.2\%         & 5   & 738          & 13341          & 73.0\%        & 99     & 600     & 1103           \\
lady             & 73.2\%         & 5   & 696          & 12             & 75.0\%        & 99     &600      &6                 \\         \hline
{\bf Average}   &{\bf 79.2\%}     &-    &980         &{\bf 14897.9}   &{\bf 85.7}\%   &99      &600      &{\bf 1054.2}     \\ \hline
\end{tabular}
\caption{Comparison of event sequence generation algorithms, where coverage (i.e. CRG) \emph{only} includes that of event sequence execution.\label{Table:comgen}}
\vspace{-2em}
\end{table}

In this study, we answer {\bf RQ4} by performing two experiments and comparing quality of event sequences generated by the long event sequence generation (Algorithm~\ref{alg:randomltc})
and baseline event sequence generation with POR (Algorithm~\ref{alg:baseline}).
In the first experiment, \tool constructs the FSM model by running Algorithm~\ref{alg:mc} two times using our coarse-grained state abstraction and the random event selection strategy, and generates two event sequences from
the FSM model, while the maximum bounds (i.e., Max. Length) are 9, 39, 69, 99, 129 and 159.
In the second experiment, we first generate an FSM model using the same configuration with a fixed maximum bound 99. For each application, from the same FSM model, \tool with Algorithm~\ref{alg:randomltc} enabled iteratively generates and executes event sequences with maximum bound 99 within 600s time bound.
Meanwhile, \tool with Algorithm~\ref{alg:baseline} enabled generates and executes all event sequences with redundant sequences pruned up to some maximum bound so that the execution time exceeds 600s,
moreover, the execution terminates when the execution time reaches 1200s.

Figure~\ref{Figure:sd} shows the results of the first experiment, where the X-axis is the maximum bound, and Y-axis is achieved coverage with that bound, and the black bold line shows the trend of average coverage.

Overall, the average coverage increases quickly when the maximum bound increases from 9 to 39,
but only slightly when it increases from 39 to 159.

Table~\ref{Table:comgen} shows the results of the second experiment,
where 
Columns 2-5 (resp. Columns 6-9) show the coverage (of testcase execution \emph{only}), maximum sequence length, execution time and number of event sequences after running \tool
with Algorithm~\ref{alg:baseline} (resp. Algorithm~\ref{alg:randomltc}) enabled.
Overall, the average coverage of Algorithm~\ref{alg:randomltc} (i.e., long event sequence generation)  is  6.5\% higher than that of  Algorithm~\ref{alg:baseline}
(i.e., baseline event sequence generation with POR) with less execution time.
In particular, the coverage improvement of Algorithm~\ref{alg:randomltc} is more prominent for applications \emph{case2}-\emph{case4}, \emph{chinabox}, \emph{3dmodel} and \emph{cubuild}. These results also confirm that
executing fewer long  event sequences may achieve higher coverage than executing more short event sequences.

\subsection{Discussions}
The coverage improvement, as shown in the experiment, is not significant using the weighted event selection.
There are two possible reasons. First, weighted event selection relies upon DOM event dependencies which are
computed by {\sc JSDep}. {\sc JSDep} uses context, path, flow and object-insensitive static analysis, hence
the dependencies may be not sufficiently fine-grained. Using more accurate DOM event dependencies may improve the effectiveness of the weighted event selection. Second, the selected benchmarks might not be representative of real-world JavaScript programs that are DOM event dependency intensive. \tool currently supports off-line testing (i.e. when source code is available) when DOM event dependency is enabled. Our approach is also applicable in on-line testing if the DOM event dependency is computed dynamically, as it does not
need to record and replay.

We note some limitations of the experiments. The experiments are based on a benchmark that includes only 4 large-scale Web applications (with more than 1k LOC, maximum 5k LOC).
More experiments are needed to assess randomized algorithms, for instance, the statistical significance of the coverage improvement~\cite{AB14}.

\section{Related Work}\label{sec:rel}
We discuss the related work  in the areas of model-based testing and automated JavaScript Web application testing.
%

\subsection{Model-based Automated Testing}
Model-based testing (MBT) has been widely used in software testing (cf. \cite{DT10,UPL12,LLS18} for surveys). Mainstream
MBT techniques differ mainly in three aspects: models of the software under test, model construction, and testcase generation.
Several models, such as state-based (e.g., pre-/post-condition)
and transition-based (e.g., UML and I/O automata), have been proposed. 
The FSM model used in this work is one of the transition-based models.
Model construction is one of the most important tasks in MBT.
It is usually time-consuming and error-prone to manually construct models for GUI-based applications~\cite{JPM13,TKH11}. Therefore, most works use static/dynamic analysis to construct models, for example, \cite{Amalfitano2015MobiGUITAR,Yang2013A,Baek2016Automated,Su2017stoat,CA15,XM06,Arc12,RHZ18a} for mobile/GUI applications.
However, it is rather difficult to statically construct models for JavaScript
Web applications due to their dynamic characteristics~\cite{Mesbah2012Crawling}.
Regarding works on model construction for JavaScript
Web applications~\cite{RT01,AOA05,MBD08,Marchetto2009State,Mesbah2012Crawling,Dallmeier2012WebMate};
\cite{RT01,Marchetto2009State,AOA05} have to construct a model manually. \cite{RT01} extracts the model via static analysis, 
but lacks of considering dynamic nature of JavaScript;
\cite{MBD08,Mesbah2012Crawling} construct FSM models via dynamic analysis to crawl Web applications.
The main difference between our work and theirs~\cite{MBD08,Mesbah2012Crawling} is the way in which the model is constructed. Our model construction pursues larger depth without backtracking, but does not cover all possible event sequences,
whereas \cite{MBD08,Mesbah2012Crawling} cover all the possible event sequences up to a length bound with backtracking.
\cite{Dallmeier2012WebMate} also reported to construct FSM models, but did not give detail of their algorithm, nor included JavaScript coverage.
Existing testcase generation algorithms mainly focus on the systematic generation of event sequences with a rather limited length bound due to ``test sequence explosion" problem.
Our approach generates long event sequences, but strategically avoid covering all possible event sequences (up to a length bound) to mitigate the exponential blowup problem. The tool {\sc Stoat}~\cite{Su2017stoat} considered model-based testing for Android apps, but with different testcase generation strategy as ours.

\subsection{JavaScript Web Application Automated Testing}
Web application testing has been widely studied in the past decade, differing mainly in targeted
Web programming languages (e.g., PHP~\cite{AH11,WYCDIS08,AKDTDPE08} and JavaScript~\cite{artzi2011framework,Mesbah2012Crawling,HT12}),
and testing techniques (e.g., model-based testing~\cite{DT10,UPL12,LLS18}, mutation testing~\cite{MP15,Mirshokraie2013Efficient,Marchetto2009State}, search-based testing~\cite{AH11,GFZ12,TGZ14,Zeller17}
and symbolic/concolic testing~\cite{SAHMMS10,Gibbs2013Jalangi,li2014symjs}; cf. \cite{LDD14,AGMPSSS17} for surveys). We mainly compare with works on automated testing of JavaScript Web Application.

Test sequences of JavaScript Web Applications consist of event sequences and input data for each event.
Existing works create event sequences via exploring the state space by randomly selecting events  with heuristic search strategies~\cite{SAHMMS10,artzi2011framework,Pradel2014EventBreak,Mesbah2012Crawling,Mirshokraie14}.
For instance, Kudzu~\cite{SAHMMS10} and {\sc Crawljax}~\cite{Mesbah2012Crawling} randomly select available events.
Moreover, {\sc Crawljax} relies on a heuristic approach for detecting event handlers, hence may not be able to detect all of them. {\sc Artemis}~\cite{artzi2011framework} uses the heuristic strategy based on the observed read and write operations by each event handler in an attempt to exclude sequences of non-interacting event handler executions.
EventBreak~\cite{Pradel2014EventBreak} uses the heuristic strategy based on performance costs in terms of the number of conditions in event handlers
in an attempt to analyze responsiveness of the application.
These approaches usually cover all the sequences up to a given, usually small length bound. 
In order to explore long event sequences in limited time,
delta-debugging based method~\cite{YPGS15} and partial-order reduction~\cite{sung2016static}
were proposed for pruning redundant event sequences.
Our approach does not cover all possible event sequences, hence can create long event sequences 
within the time budget. Experimental results show that
our approach can achieve a high line coverage than {\sc Artemis} even with partial-order reductions~\cite{sung2016static}.

Another research line in automated testing of JavaScript Web application is to generate high quality input data of events using symbolic/concolic testing, e.g.,
\cite{SAHMMS10,Gibbs2013Jalangi,li2014symjs,SNGC15,TULG15,DRS16}.
These approaches are able to achieve high coverage, but
heavily rely on the underlying constraint solver. They generally do not scale well
for large realistic programs, because the number of feasible execution paths of a program often increases exponentially 
in the length of the path. Our work focuses on the generation of long event sequences, but choose the input data randomly, which is orthogonal, and could be complementary, to the more advanced input data generation methods.
A transfer technique has been proposed 
based on the automation engine framework {\sc Selenium} 
 to transfer tests from one JavaScript Web application to another~\cite{RHZ18b}. This is orthogonal to this work.

\section{Conclusion and Future Work}\label{sec:conc}
We have proposed a model-based automated testing approach for JavaScript Web applications.
Our approach distinguishes from others in making use of \emph{long} event sequences in both FSM model construction and testcase generation from the FSM model.
%
We have implemented our approach in a tool \tool and evaluated it on  a collection of benchmarks. 
The experimental results showed that our new approach is more efficient and effective than {\sc Artemis} and {\sc JSdep}.
Furthermore, we empirically found that longer test sequences can achieve a high line coverage.


For future work, we plan to experiment on more larger benchmarks and to have a thorough statistical analysis. In particular, the current evaluation is based on coverage, but it would also be interesting to evaluate the fault detection. Furthermore, we note that there are several ways to obtain long sequences based on FSM, for instance, by following the dependency chains or giving priority to the novel states. These strategies deserve further exploration.

%
%
%
%



\bibliographystyle{IEEEtran}

\end{document}